\documentclass[12pt]{article}
\usepackage{epsfig}

\usepackage{amsmath}
\usepackage{latexsym}

\def\lsim{\mathrel{\rlap{\raise 2.5pt \hbox{$<$}}\lower 2.5pt
\hbox{$\sim$}}}
\def\gsim{\mathrel{\rlap{\raise 2.5pt \hbox{$>$}}\lower 2.5pt
\hbox{$\sim$}}}
\def\thW{\theta_{\rm W}}
\def\GeV{{\rm GeV}}
\def\TeV{{\rm TeV}}
\def\dd{{\rm d}}

\def \sup{^{\vphantom{2}}}

\catcode`@=11
\def\citer{\@ifnextchar [{\@tempswatrue\@citexr}{\@tempswafalse\@citexr[]}}
 
\def\@citexr[#1]#2{\if@filesw\immediate\write\@auxout{\string\citation{#2}}\fi
  \def\@citea{}\@cite{\@for\@citeb:=#2\do
    {\@citea\def\@citea{--\penalty\@m}\@ifundefined
       {b@\@citeb}{{\bf ?}\@warning
       {Citation `\@citeb' on page \thepage \space undefined}}%
\hbox{\csname b@\@citeb\endcsname}}}{#1}}
\catcode`@=12

\newenvironment{capt}{
\phantom{mmmm}
\vspace*{10mm}
\parindent=0pt
\addtocounter{figure}{1}

\begin{minipage}[t]{160mm}
\small\sl Figure~\thefigure.\ }{\end{minipage}
\vspace*{-5mm}}

\begin{document}

\thispagestyle{empty}

\begin{flushright}
{\tt University of Bergen, Department of Physics}    \\[2mm]
{\tt Scientific/Technical Report No.~1999-09}    \\[4mm]
{\tt ISSN 0803-2696} \\[3mm]
{hep-ph/9911295} \\[3mm]
{November 1999}       
\end{flushright}

\vspace*{44mm}

\begin{center}
{\bf \large
Multiple Higgs Production and Measurement of \\[5mm]
Higgs Trilinear Couplings in the MSSM}
\vspace{10mm}

P.~Osland $^a$ and P. N. Pandita$^b$

\vspace{3mm}

{\em $^a$ Department of Physics, University of Bergen, 
All\'{e}gaten 55, N-5007 Bergen, Norway\\
$^b$ Department of Physics, North Eastern Hill University,
Shillong 793 022, India}

\end{center}
\newpage
\phantom{AAA}
\clearpage
\pagenumbering{arabic}

\begin{center}
{\bf{\Large Multiple Higgs Production and Measurement of \\[2mm]
Higgs Trilinear Couplings in the MSSM}}
\footnote{To appear in the
{\it Proceedings of XIVth International Workshop: High Energy
Physics and Quantum Field Theory}, Moscow, Russia,
27 May --  2 June 1999.}
\vspace{4mm}

P. Osland$^a$  and P. N. Pandita$^b$\\
\vspace{3mm}
$^a$Department of Physics, University of Bergen, 
N-5007 Bergen, Norway\\
$^b$ Department of Physics, North Eastern Hill University, \\
Shillong 793 022, India

\end{center}

\begin{abstract}
An elementary Higgs boson, which is the  
remnant of the spontaneous symmetry breaking 
mechanism in the Standard Model, remains one of the most elusive particles. 
If a  Higgs boson candidate is  found at future accelerators,
it becomes necessary to determine its properties,
beyond the mass, production cross section and decay rates.
The other crucial properties relate to the self-couplings of the Higgs boson, 
which are necessary to reconstruct the Higgs potential, 
and thereby confirm the mechanism of spontaneous symmetry breaking.
In this paper we review the question of the measurability of some of 
the trilinear couplings of the neutral Higgs bosons of the 
Minimal Supersymmetric Standard Model at a high-energy $e^+e^-$ collider.
\end{abstract}

  
\section{Introduction}

The Higgs mechanism, which is responsible for the generation of all
particle masses in the Standard Model~(SM), is as 
yet an experimentally untested mechanism. This mechanism of mass generation
leaves behind a remnant in the form of  scalar particle(s),
the Higgs boson(s), which has eluded the experimental search so far.
The Higgs particle is expected to be discovered at the LHC,
if not  at LEP2 \cite{Zerwas}.
Current estimates from precision electroweak data \cite{LEP} 
suggest that it is rather light. 
Although a light Higgs particle would be consistent with the Standard
Model, it is natural only in  a supersymmetric framework~\cite{HPN}.
The minimal version of the Supersymmetric 
Standard Model (MSSM) contains two Higgs doublets $(H_1, H_2)$ 
with opposite hypercharges~\cite{GHKD}: 
$Y(H_1) = -1$, $Y(H_2) = +1$, so as to 
generate masses for up- and down-type
quarks (and leptons), and to cancel triangle gauge anomalies. 
After spontaneous symmetry breaking induced by the neutral 
components of $H_1$ and $H_2$ obtaining vacuum 
expectation values, $\langle H_1\rangle = v_1$, 
$\langle H_2\rangle = v_2$, $\tan\beta = v_2/v_1$, 
the MSSM contains two neutral $CP$-even ($h$, $H$), one neutral 
$CP$-odd ($A$), and two charged ($H^{\pm}$) Higgs bosons. 
Because of gauge invariance and supersymmetry,
all the Higgs masses
and the Higgs couplings in the MSSM can be described (at tree level) 
in terms of only  two parameters, which are usually chosen to be
$\tan\beta$ and $m_A$, the mass of the $CP$-odd Higgs boson.
Once a light Higgs boson is discovered, a detailed measurement of 
its branching ratios should, in principle, enable one 
to distinguish between a SM Higgs boson and the lightest 
MSSM Higgs boson.

Apart from the large number of Higgs bosons in the minimal
version of the supersymmetric standard model,
there is more to the MSSM Higgs sector than the branching ratios.
For a complete analysis of the Higgs sector, one should also
measure the trilinear and quartic
self-couplings of the Higgs boson, 
which in the MSSM are determined (at the tree level) 
by the gauge couplings.
The measurability of trilinear 
couplings involving the light Higgs  boson
was investigated by Djouadi, Haber and Zerwas \cite{DHZ}.
In this preliminary study, 
it was concluded that the trilinear couplings $\lambda_{Hhh}$ 
and $\lambda_{hhh}$, where $h$ and $H$ denote the two neutral, 
$CP$-even Higgs bosons, could be measured at a 
high-energy $e^+ e^-$ linear collider.
A more detailed study, including the squark mixing, 
of the measurability of the
trilinear Higgs couplings was 
carried out in \cite{OP98}.
Recently, the leading  two-loop effects have also been incorporated
in the calculations \cite{Muhlleitner}.

All the trilinear self-couplings of the physical Higgs particles 
can be predicted theoretically~(at the tree level) in terms of $m_A$ 
and $\tan\beta$.  Once a light Higgs boson is discovered, the
measurement of these trilinear couplings  can be used to reconstruct
the Higgs potential of the MSSM. This will go a long way in
establishing the Higgs mechanism as the basic mechanism of 
spontaneous symmetry breaking in gauge theories. 

We have considered
in detail~\cite{OP98} the question of the possible measurements of
some of the trilinear Higgs couplings of the MSSM 
at a high-energy $e^+ e^-$ 
linear collider that will operate at an energy of 500~GeV 
with an integrated luminosity per year of  
${\mathcal L}_{\rm int} = 500~\mbox{fb}^{-1}$ \cite{NLC}.
We focus on the trilinear Higgs couplings 
$\lambda_{Hhh}$ and  $\lambda_{hhh}$, 
involving the $CP$-even Higgs bosons.
They are rather small with respect to the corresponding trilinear coupling
$\lambda_{hhh}^{\rm SM}$ in the SM (for a given mass 
of the lightest Higgs boson $m_h$), 
unless $m_h$ is close to the upper value (the decoupling limit).

\section{Trilinear Higgs couplings}
In units of $gm_Z/(2\cos\thW)=(\sqrt{2}G_F)^{1/2}m_Z^2$,
the {\it tree-level} trilinear Higgs couplings of the MSSM,
that we shall discuss, are
given by \cite{GHKD}:
\begin{eqnarray} 
\lambda_{hhh}^0 & = & 3 \cos2\alpha \sin(\beta + \alpha), 
\label{Eq:lambda-hhh0} \\
\lambda_{Hhh}^0 & = & 2\sin2\alpha \sin(\beta + \alpha) - \cos 2\alpha
\cos(\beta + \alpha),
\label{Eq:lambda-Hhh0}
\end{eqnarray}
where $\alpha$ is the mixing angle in the $CP$-even Higgs sector, which 
is determined by the parameters of the $CP$-even
Higgs boson mass matrix. 

The trilinear Higgs couplings 
$\lambda_{Hhh}$ and $\lambda_{hhh}$ of the neutral Higgs
bosons in the Minimal Supersymmetric Standard Model~(MSSM),
Eqs.~(\ref{Eq:lambda-hhh0}) and (\ref{Eq:lambda-Hhh0}),
involve the $CP$-even Higgs bosons $h$ and $H$.
These  trilinear couplings can be measured through
the multiple production of the Higgs bosons
at  high-energy $e^+  e^-$ colliders. 
The relevant production mechanisms that we shall  
consider are the production of 
the heavier $CP$-even Higgs boson via
$e^+e^- \rightarrow ZH$, in association with 
the $CP$-odd Higgs boson ($A$) in $e^+e^- \rightarrow AH$, 
or via the fusion process $e^+e^- \rightarrow \nu_e \bar\nu_e H$, 
with $H$ subsequently decaying through $H \rightarrow hh$.
The multiple production
of the light Higgs boson through Higgs-strahlung of $H$, and 
through production of $H$ in association with the $CP$-odd Higgs 
boson can be used to 
extract the trilinear Higgs coupling $\lambda_{Hhh}$.
The non-resonant fusion mechanism for multiple
$h$ production, $e^+e^-\to \nu_e\bar\nu_e hh$, involves
two trilinear Higgs couplings, $\lambda_{Hhh}$ and $\lambda_{hhh}$,
and is useful for extracting $\lambda_{hhh}$.
For the extraction of other MSSM trilinear Higgs couplings, 
$\lambda_{HHh}$, 
$\lambda_{HHH}$, 
$\lambda_{hAA}$, and  
$\lambda_{HAA}$, 
at $e^+ e^-$ colliders, see~\cite{Muhlleitner}.

At the tree level, the CP-even Higgs boson $h$ is rather light,
and $m_h \le m_Z$ holds. However, there are large radiative corrections
to this result and a new bound is set, $m_h \le 135$ GeV.
The dominant one-loop radiative corrections are proportional
to $(m_t/m_W)^4$, and to a set of complicated 
functions depending on the squark masses \cite{ERZ1,BBSP}.
Recently, the dominant two-loop radiative corrections~\cite{twoloop}
to the Higgs sector have been evaluated. The two-loop 
contributions yield a large correction to the one-loop result,
and as a consequence $m_h$ is reduced by up to $\sim 20$ GeV,
which is particularly important for low values of $\tan\beta.$
In Fig.~\ref{Fig:mhl} we plot the two-loop corrected mass of the lightest 
Higgs boson as a function of $m_A$ and $\tan\beta$ for two values of
the mixing parameters $A$ and $\mu$, as indicated\footnote{The LEP 
experiments have obtained strong
lower bounds on the mass of the lightest Higgs boson, and are
beginning to rule out significant parts of the small-$\tan\beta$ 
parameter space.
ALEPH finds a lower limit of $m_h>72.2$~GeV, irrespective of $\tan\beta$,
and a limit of $\sim 88$~GeV for $1<\tan\beta\lsim2$ \cite{ALEPH98}.}.

We shall include one-loop radiative corrections
\cite{ERZ1,BBSP}, as well as the leading
two-loop corrections \cite{twoloop}, to the Higgs sector in 
our calculations.
In particular, we take into account \cite{OP98}
the parameters $A$ and $\mu$, the soft supersymmetry
breaking trilinear parameter and the bilinear Higgs(ino)  
parameter in the superpotential.
These parameters determine the stop masses,
\begin{equation}
m_{\tilde t_{1,2}}^2   =  m_t^2 + \tilde m^2 \pm 
m_t(A + \mu\cot\beta)
\end{equation}
which enter through the radiative corrections to the Higgs masses
as well as to the Higgs trilinear couplings.

\begin{figure}[htb]
\refstepcounter{figure}
\label{Fig:mhl}
\addtocounter{figure}{-1}
\begin{center}
\setlength{\unitlength}{1cm}
\begin{picture}(12,7.6)
\put(0,1)
{\mbox{\epsfysize=7.5cm\epsffile{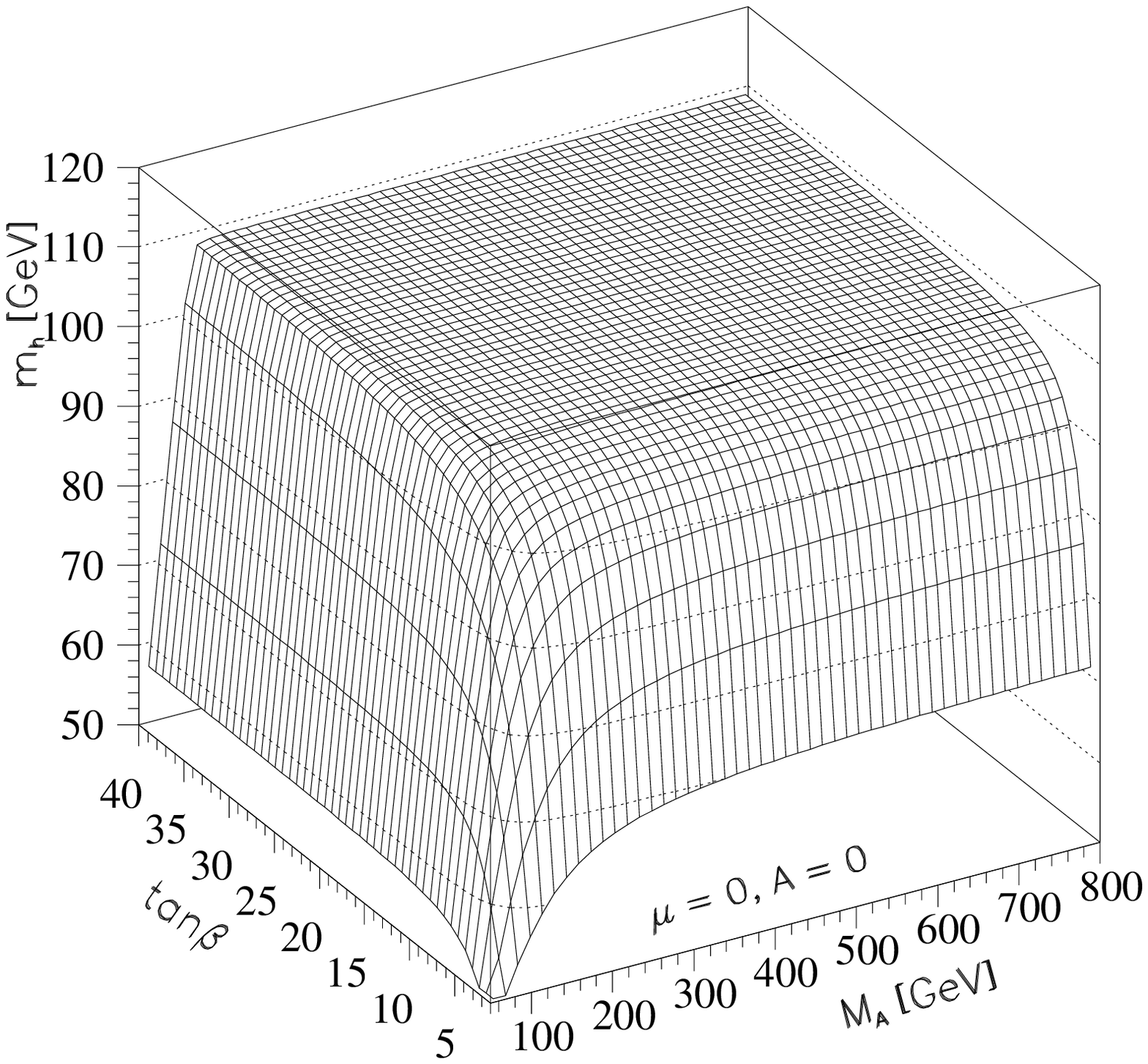}}
 \mbox{\epsfysize=7.5cm\epsffile{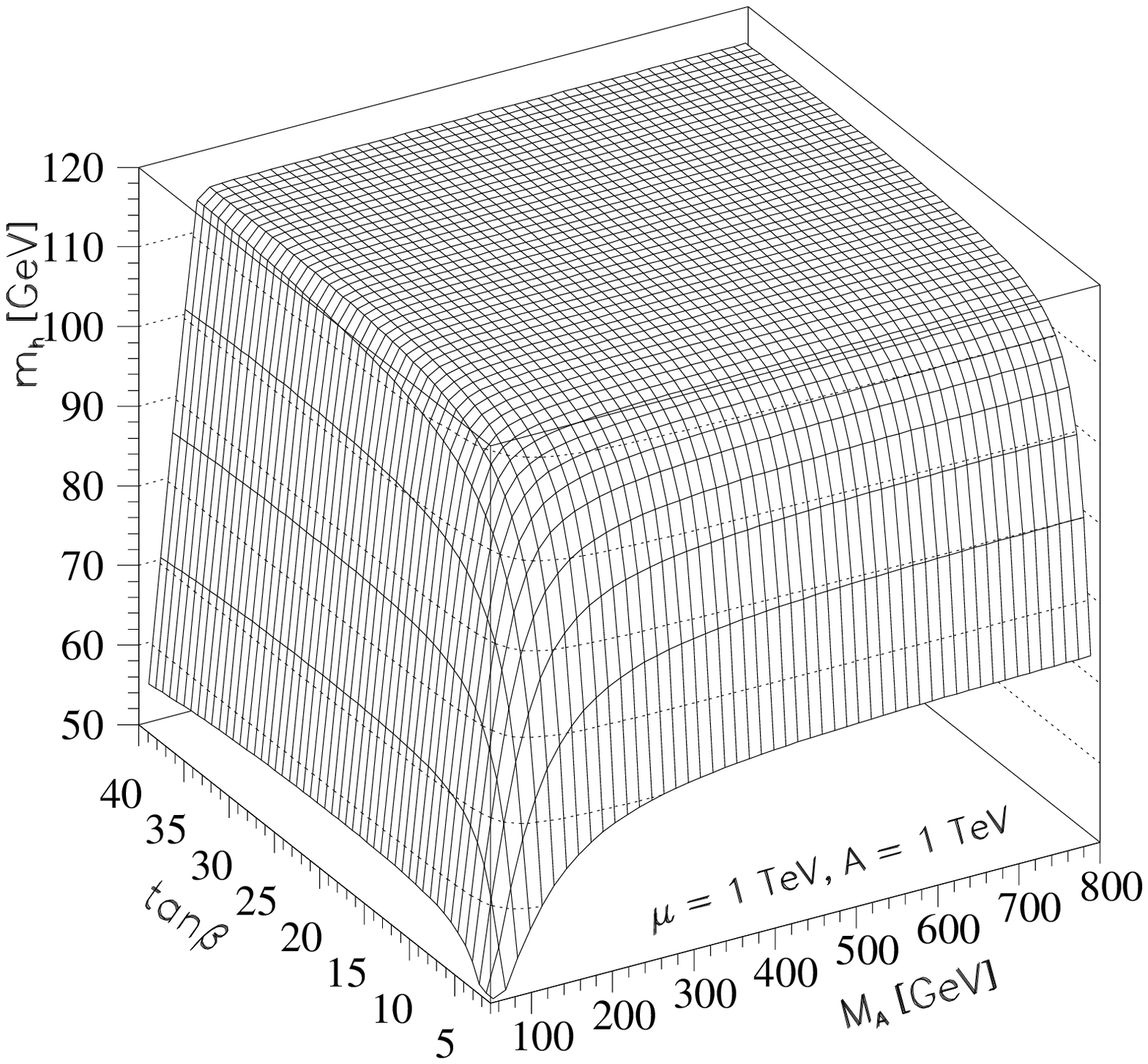}}}
\end{picture}
\vspace*{-22mm}
\begin{capt}
Mass of the lightest $CP$-even Higgs boson, $m_h$
{\it vs.} $m_A$ and $\tan\beta$.
Two cases are considered for the mixing parameters $\mu$ and $A$,
as indicated.
Two-loop corrections are included \cite{twoloop}.
\end{capt}
\end{center}
\end{figure}

The trilinear couplings depend significantly on $m_A$,
and thus also on $m_h$. This is shown in Fig.~\ref{Fig:lam-mh},
where we compare $\lambda_{Hhh}$, $\lambda_{hhh}$ and $\lambda_{hAA}$
for two different values of $\tan\beta$.
For a given value of $m_h$, the values of these
couplings significantly depend on the soft supersymmetry-breaking 
trilinear parameter $A$, as well as on $\mu$.
\begin{figure}[htb]
\refstepcounter{figure}
\label{Fig:lam-mh}
\addtocounter{figure}{-1}
\begin{center}
\setlength{\unitlength}{1cm}
\begin{picture}(12,7.5)
\put(0.5,1)
{\mbox{\epsfysize=7.0cm\epsffile{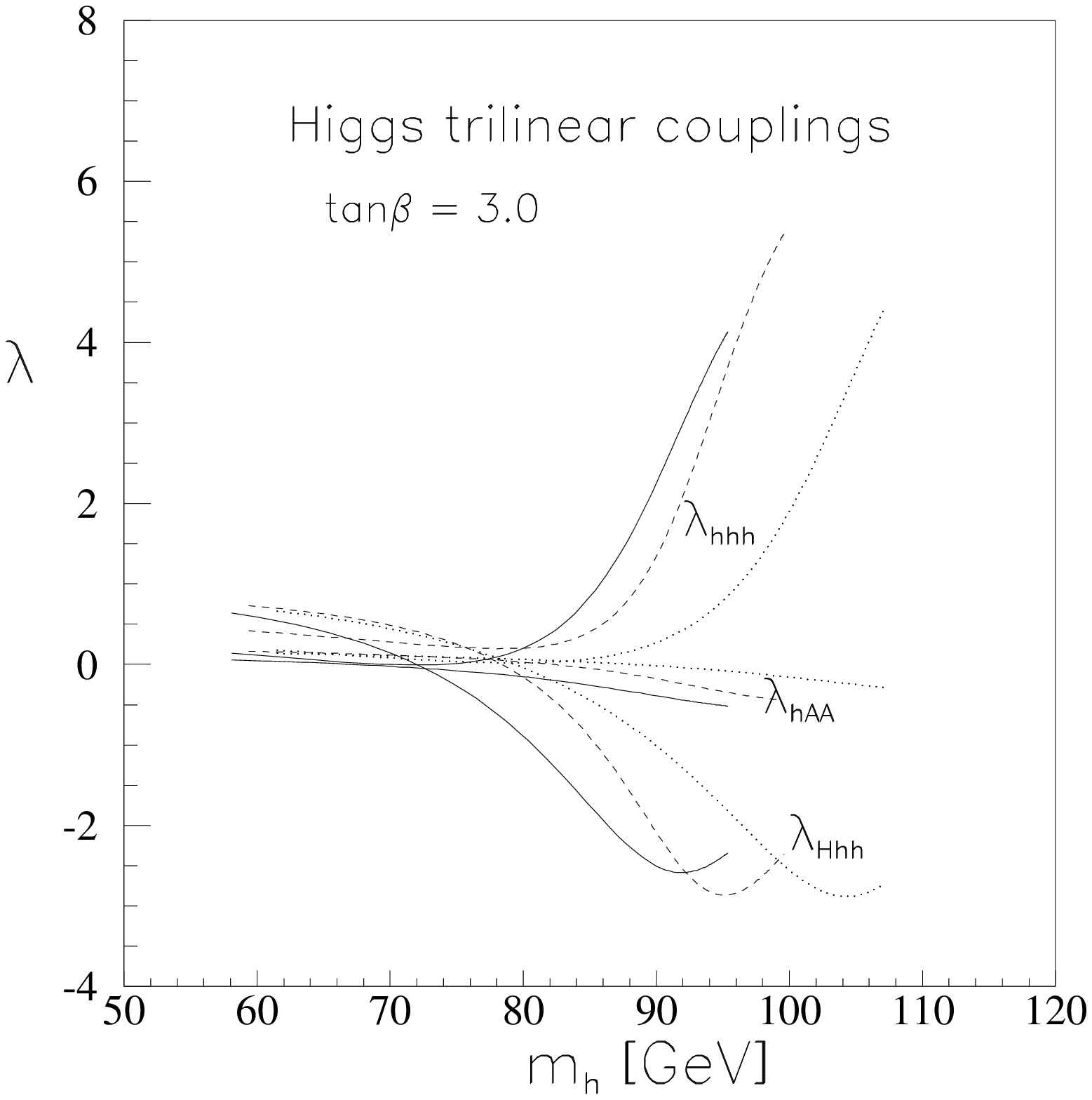}}
\hspace*{-5mm}
\mbox{\epsfysize=7.0cm\epsffile{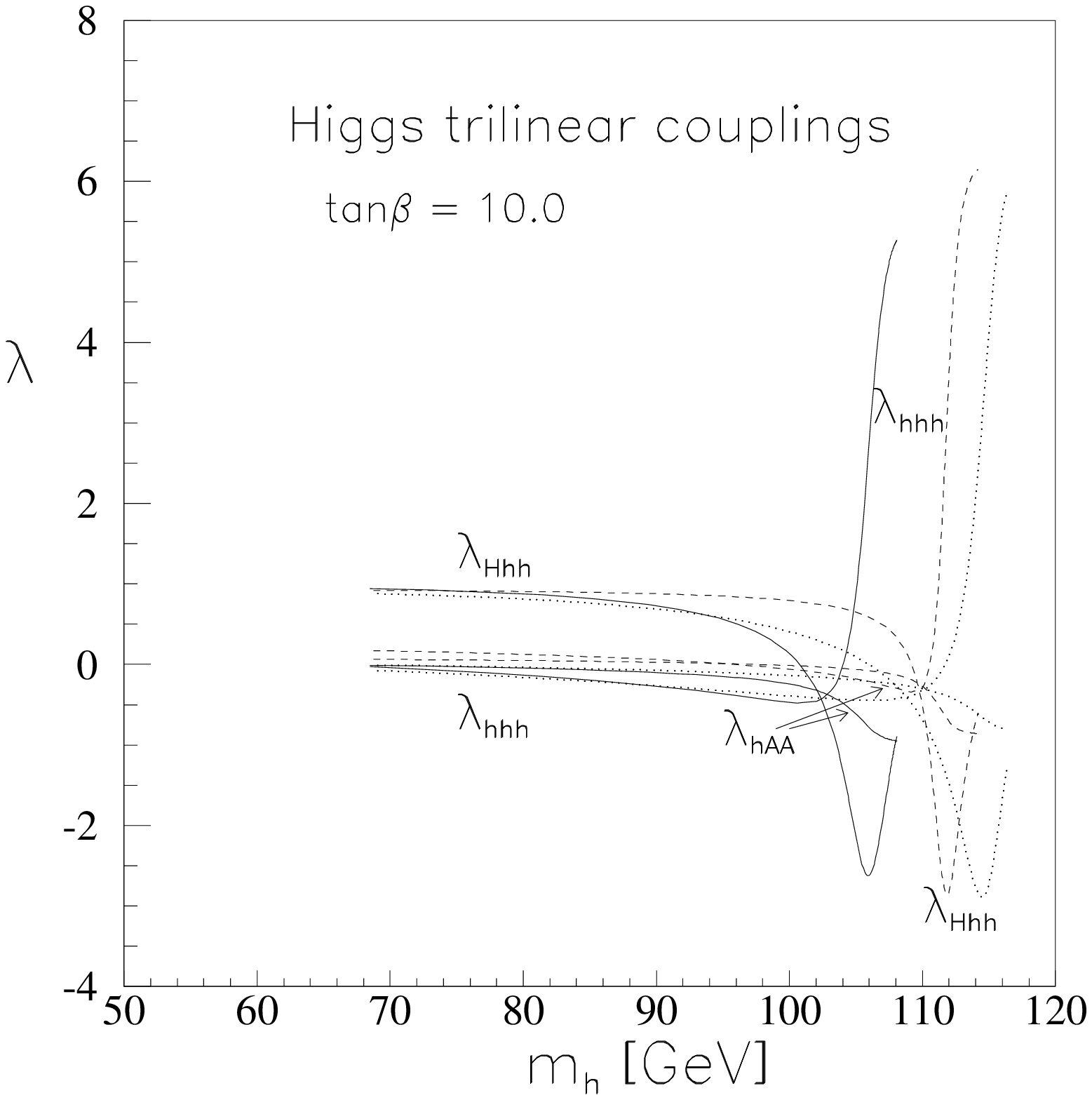}}}
\end{picture}
\vspace*{-18mm}
\begin{capt}
Trilinear Higgs couplings $\lambda_{Hhh}$, $\lambda_{hhh}$ and
$\lambda_{hAA}$ as functions of $m_h$ for 
$\tan\beta=3.0$ and $\tan\beta=10.0$.
Each coupling is shown for $\tilde m=1~\TeV$, and for
three cases of the mixing parameters:
no mixing ($A=0$, $\mu=0$, solid),
mixing with $A=1$~TeV and $\mu=-1$~TeV (dotted),
as well as 
$A=1$~TeV and $\mu=1$~TeV (dashed).
\end{capt}
\end{center}
\end{figure}

As is clear from Fig.~\ref{Fig:lam-mh},
at low values of $m_h$, the MSSM trilinear couplings are rather small.
For some value of $m_h$ the couplings $\lambda_{Hhh}$ and $\lambda_{hhh}$
start to increase in magnitude, whereas $\lambda_{hAA}$ remains small.

\section{Production mechanisms}
The dominant mechanisms for the production of multiple  
$CP$-even light Higgs bosons is through the processes
\begin{eqnarray}
\left. \begin{array}{ccc}
e^+e^- & \rightarrow & ZH,AH \\ 
e^+e^- & \rightarrow & \nu_e \bar \nu_e H
\end{array}
\right\}, \qquad H \rightarrow hh, \label{Eq:res-Hhh} 
\end{eqnarray}
shown in Fig.~\ref{Fig:Feynman-resonant}. 
The heavy Higgs boson $H$ can be produced
by $H$-strahlung, in association with $A$, 
and by the resonant $WW$ fusion mechanism. 
All the diagrams of Fig.~\ref{Fig:Feynman-resonant} involve the
trilinear coupling $\lambda_{Hhh}$.

\begin{figure}[htb]
\refstepcounter{figure}
\label{Fig:Feynman-resonant}
\addtocounter{figure}{-1}
\begin{center}
\setlength{\unitlength}{1cm}
\begin{picture}(12,5.5)
\put(0.5,0.0)
{\mbox{\epsfxsize=16cm\epsffile{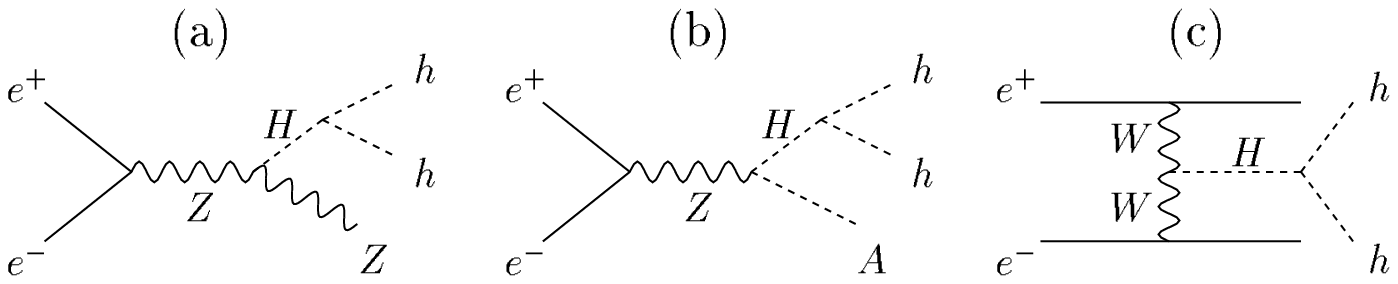}}}
\end{picture}
\vspace*{-40mm}
\begin{capt}
Feynman diagrams for the resonant production
of $hh$ final states in $e^+ e^-$ collisions.
\end{capt}
\end{center}
\end{figure}
\vspace*{-5mm}

A background to the processes (\ref{Eq:res-Hhh}) 
comes from the production of the 
pseudoscalar $A$ in association with $h$ and its subsequent decay to $hZ$
\begin{equation}
e^+e^- \rightarrow hA, \qquad A \rightarrow hZ, \label{Eq:bck-hA}
\end{equation}
leading to $Zhh$ final states.

Another  mechanism for $hh$ production is double Higgs-strahlung 
in the continuum with a $Z$ boson in the final state,
\begin{equation}
e^+e^-  \rightarrow Z^* \rightarrow Zhh. \label{Eq:Zstar}
\end{equation} 

Finally, there is also a mechanism for the  
multiple production of the lightest Higgs
boson through non-resonant $WW$ fusion in the continuum:
\begin{equation}
e^+e^-  \rightarrow \bar \nu_e \nu_e W^* W^* \rightarrow 
\bar \nu_e \nu_e hh, \label{Eq:WW-fusion}
\end{equation}
as shown in Fig.~\ref{Fig:Feynman-nonres-WW}.

\begin{figure}[hbt]
\refstepcounter{figure}
\label{Fig:Feynman-nonres-WW}
\addtocounter{figure}{-1}
\begin{center}
\setlength{\unitlength}{1cm}
\begin{picture}(12,5.0)
\put(1,0.0)
{\mbox{\epsfxsize=14cm\epsffile{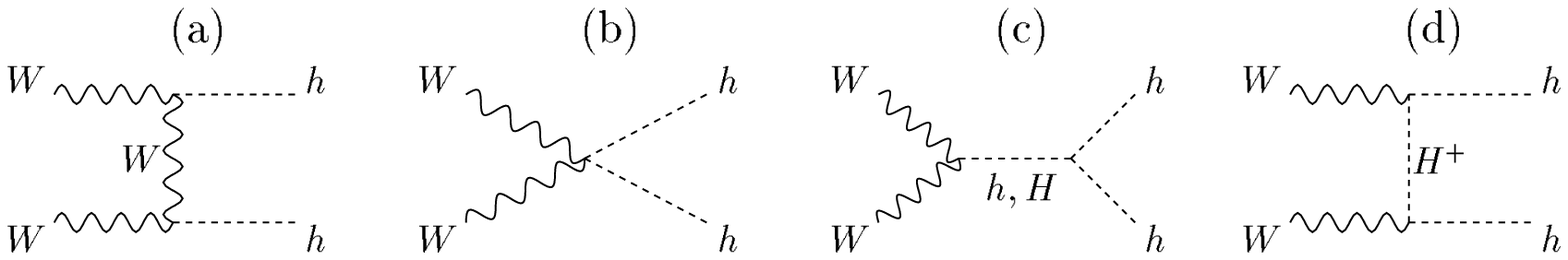}}}
\end{picture}
\vspace*{-33mm}
\begin{capt}
Feynman diagrams for the non-resonant $WW$ fusion
mechanism for the production of $hh$ states in $e^+ e^-$ collisions.
\end{capt}
\end{center}
\end{figure}

It is important to note that all the diagrams of  
Fig.~\ref{Fig:Feynman-resonant}
involve the trilinear coupling $\lambda_{Hhh}$ only. 
In contrast,
the non-resonant analogues of Figs.~\ref{Fig:Feynman-resonant}a,
\ref{Fig:Feynman-resonant}b and \ref{Fig:Feynman-resonant}c
(or \ref{Fig:Feynman-nonres-WW}c) involve both
the trilinear Higgs couplings $\lambda_{Hhh}$ and $\lambda_{hhh}$.

\subsection{Higgs-strahlung and associated production of $H$}
The dominant source for the production of
multiple light Higgs bosons in $e^+ e^-$ collisions is through 
the production of the heavier $CP$-even Higgs boson $H$ either via
Higgs-strahlung or in association with $A$,  
followed, if kinematically allowed, by the decay 
$H \rightarrow hh$.

\begin{figure}[htb]
\refstepcounter{figure}
\label{Fig:sigma-500-1500}
\addtocounter{figure}{-1}
\begin{center}
\setlength{\unitlength}{1cm}
\begin{picture}(12,7.8)
\put(0.5,1.5)
{\mbox{\epsfysize=7.0cm\epsffile{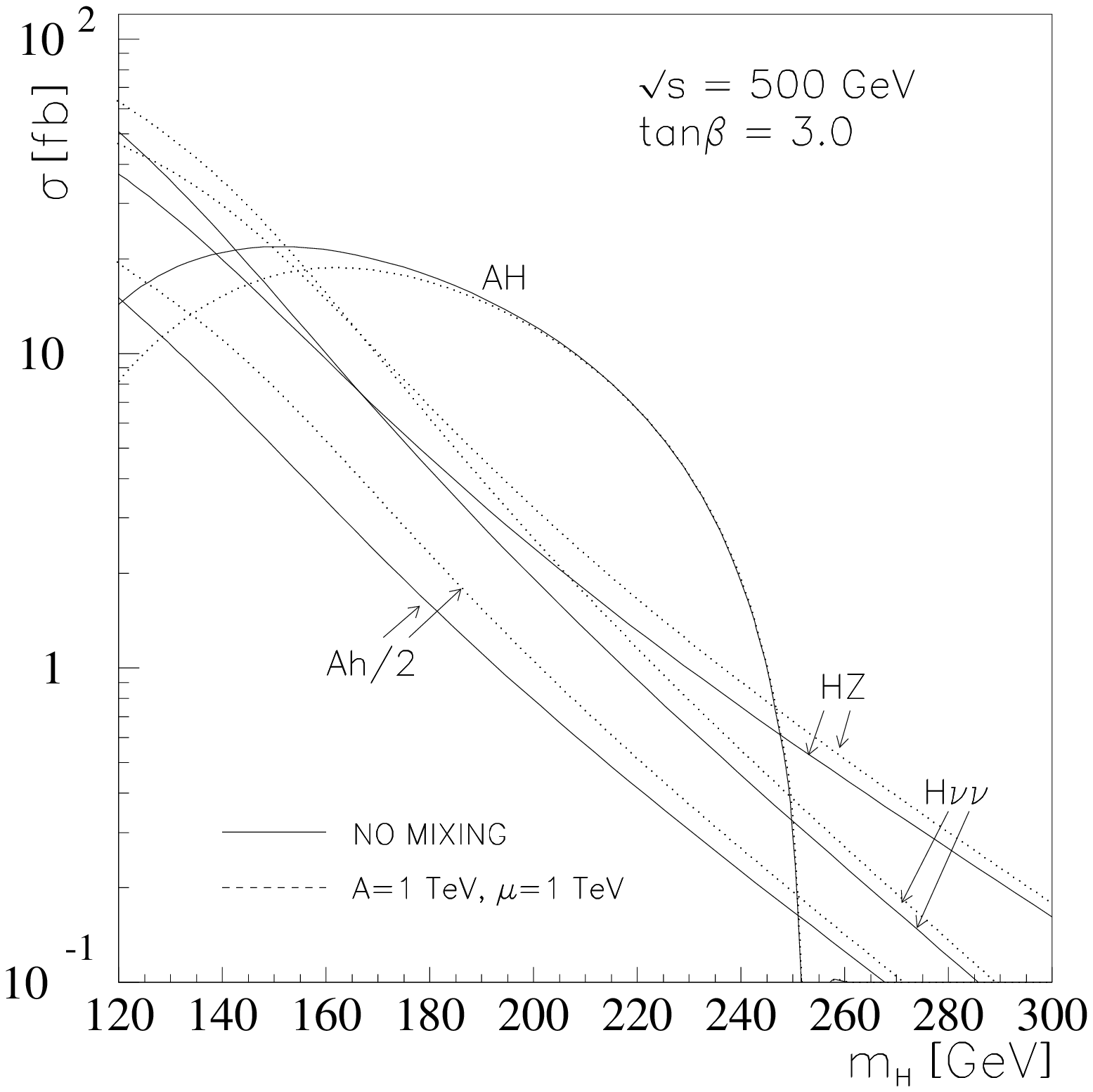}}
 \mbox{\epsfysize=7.0cm\epsffile{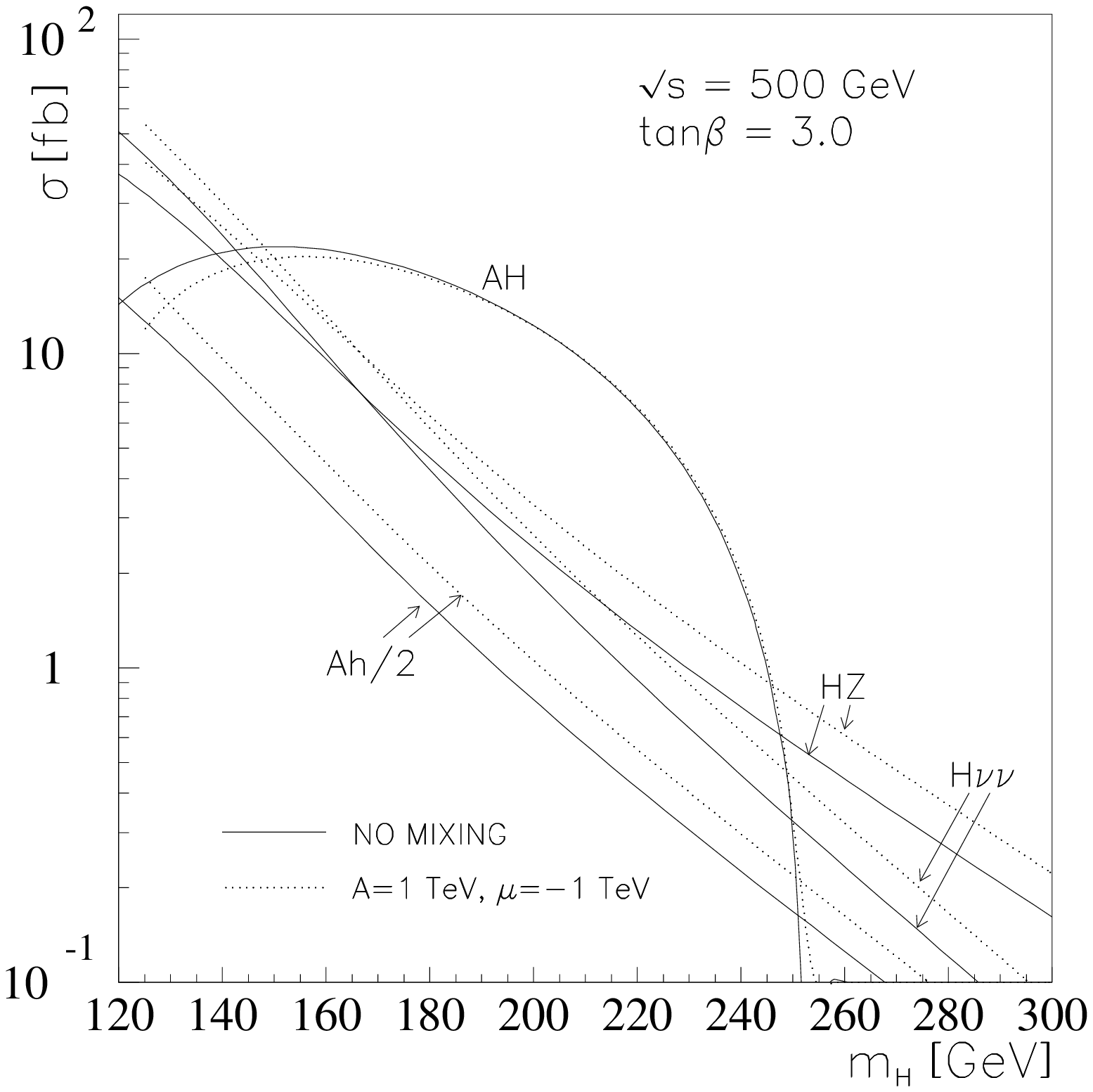}}}
\end{picture}
\vspace*{-22mm}
\begin{capt}
Cross sections for the production of the heavy Higgs
boson $H$ in $e^+ e^-$ collisions,
and for the background process in which $Ah$ is produced. 
Solid curves are for no mixing, $A=0$, $\mu=0$. 
Dashed and dotted curves refer to mixing.
\end{capt}
\end{center}
\end{figure}

In Fig.~\ref{Fig:sigma-500-1500} we plot the relevant cross sections 
\cite{PocZsi,GETAL} for the 
$e^+e^-$ centre-of-mass energy $\sqrt s = 500~\GeV$,
as functions of the Higgs mass $m_H$ and for $\tan\beta = 3.0$. 
For a fixed value of $m_H$, there is seen to be a significant
sensitivity to the squark mixing parameters $\mu$ and $A$.
We have here taken $\tilde m=1~\TeV$, a value which is adopted
throughout, except where otherwise specified.

\begin{figure}[htb]
\refstepcounter{figure}
\label{Fig:BR-H-A}
\addtocounter{figure}{-1}
\begin{center}
\setlength{\unitlength}{1cm}
\begin{picture}(12,7.8)
\put(0.5,1.5)
{\mbox{\epsfysize=7.0cm\epsffile{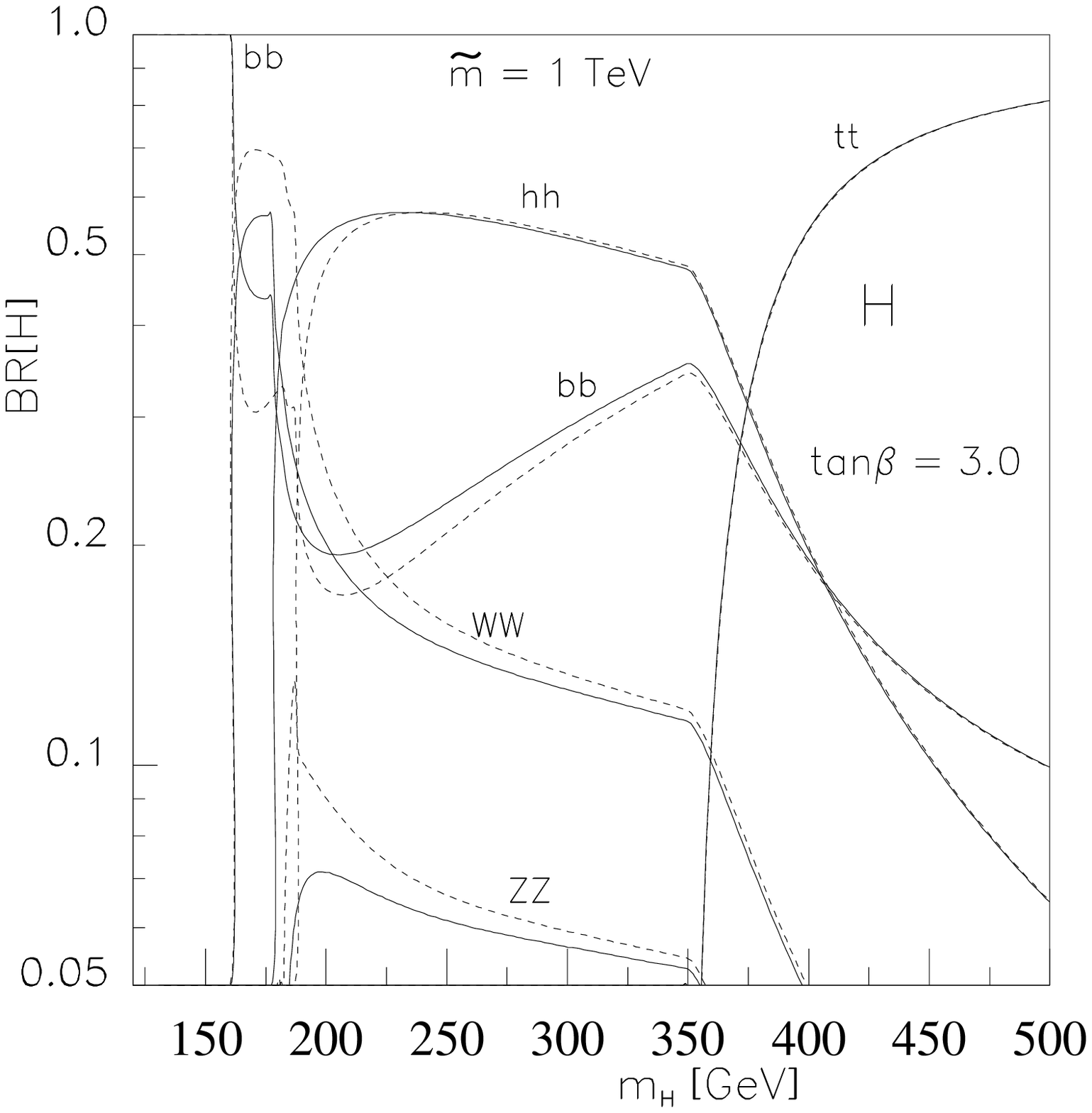}}
 \mbox{\epsfysize=7.0cm\epsffile{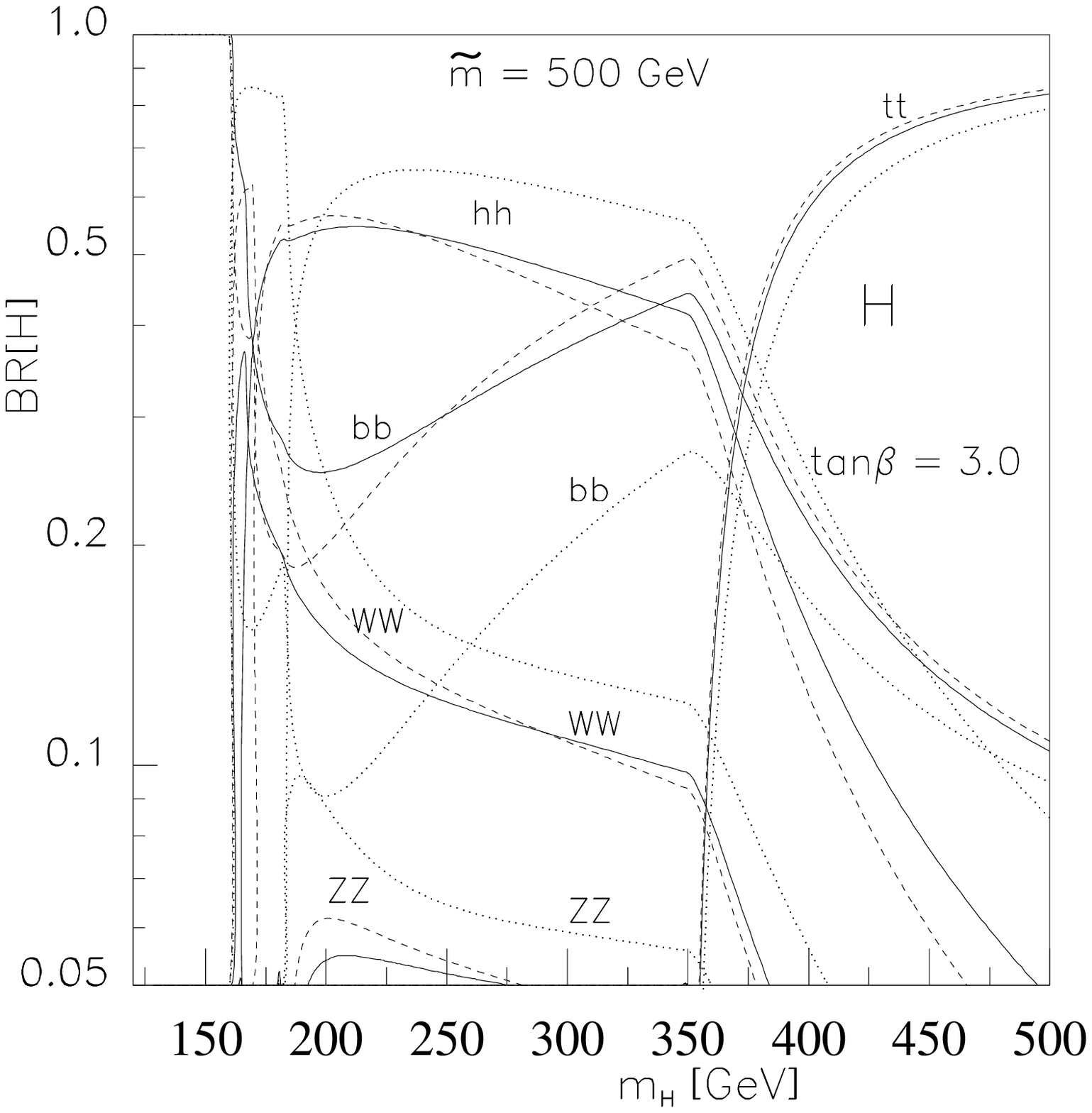}}}
\end{picture}
\vspace*{-22mm}
\begin{capt}
Branching ratios for the decay modes of the $CP$-even
heavy Higgs boson $H$, for $\tan\beta = 3.0$ and 
$\tilde m$ equal to 1~TeV or 500~GeV, 
as indicated.
Solid curves are for no mixing, $A=0$, $\mu=0$.
For $\tilde m=1$~TeV, the dashed curves refer to $A=1$~TeV and $\mu=1$~TeV,
whereas for $\tilde m=500$~GeV, the dashed (dotted) curves refer to 
$A=500$~GeV (800~GeV) and $\mu=1$~TeV (800~GeV).
\end{capt}
\end{center}
\end{figure}

A measurement of the decay rate $H \rightarrow hh$ directly
yields $\lambda_{Hhh}^2$.
But this is possible only if the decay is kinematically
allowed, and the branching ratio is sizeable (but not too close to unity).
In Fig.~\ref{Fig:BR-H-A} we show the branching ratios (at $\tan\beta=3.0$)
for the main decay modes of the heavy $CP$-even Higgs boson 
as a function of the $H$ mass \cite{DKZ1}.
Apart from the $hh$ decay mode, the other important decay modes 
are $H \rightarrow b \bar b$, $WW^*$, $ZZ^*$.
For increasing values of $\tan\beta$ (but fixed $m_h$), 
the $Hhh$ coupling gradually gets weaker (Fig.~\ref{Fig:lam-mh}),
and hence the prospects for measuring $\lambda_{Hhh}$ diminish.
Also, the decay rates can change significantly with $\tilde m$,
the over-all squark mass scale (see Fig.~\ref{Fig:BR-H-A}).

\begin{figure}[bht]
\refstepcounter{figure}
\label{Fig:hole-1000}
\addtocounter{figure}{-1}
\begin{center}
\setlength{\unitlength}{1cm}
\begin{picture}(12,7.5)
\put(0.5,1.5)
{\mbox{\epsfysize=7.0cm\epsffile{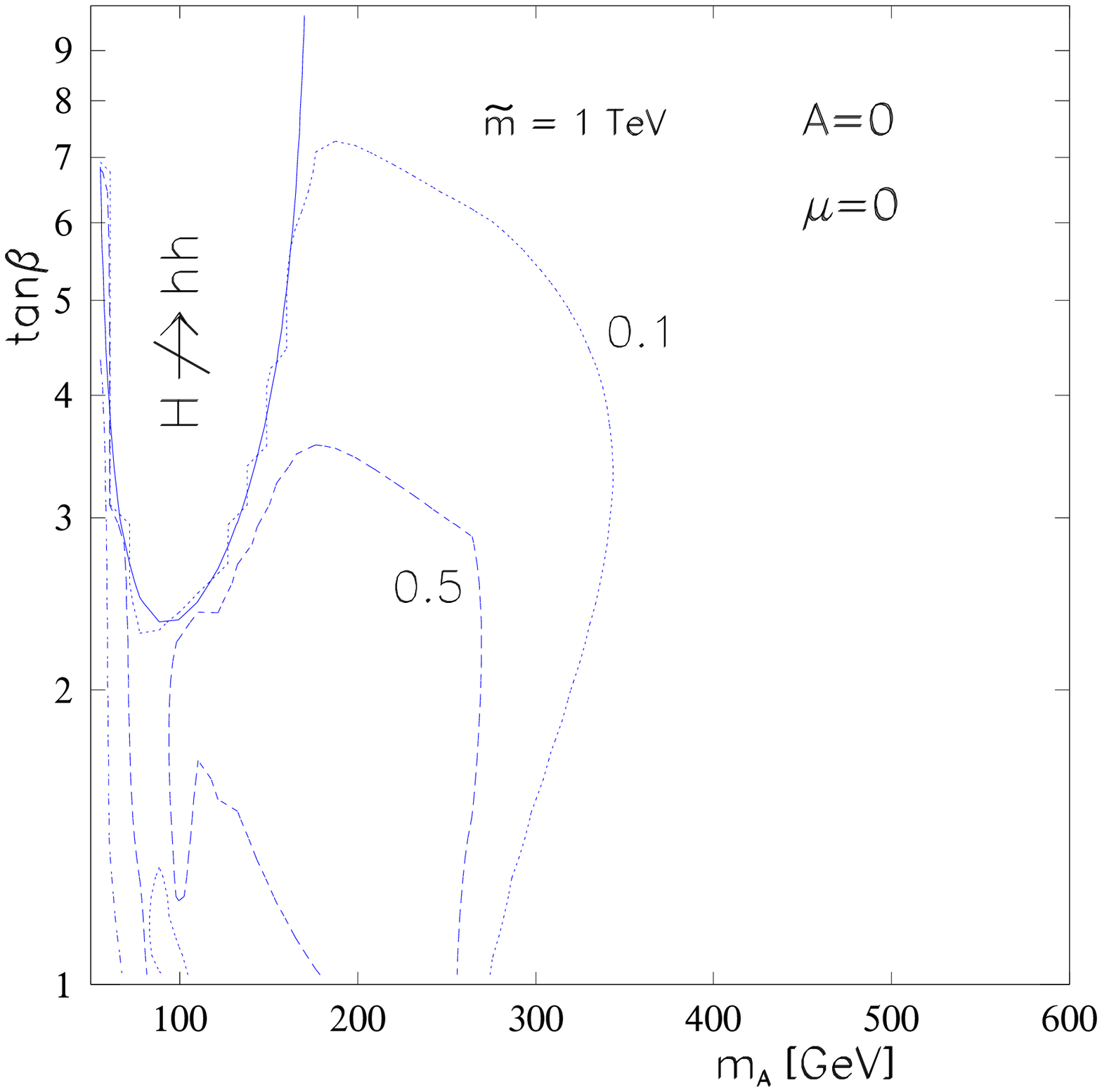}}
 \mbox{\epsfysize=7.0cm\epsffile{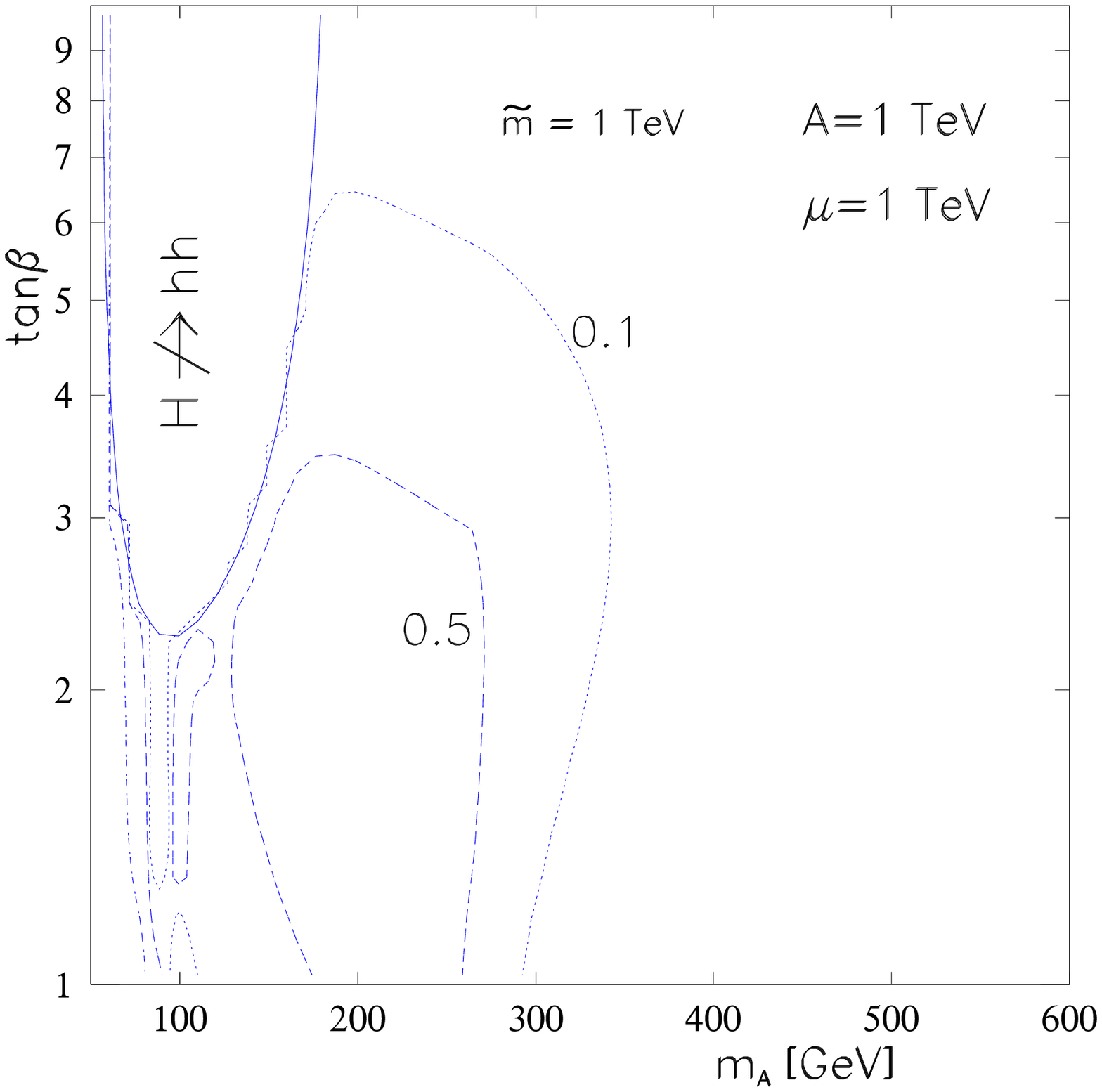}}}
\end{picture}
\vspace*{-22mm}
\begin{capt}
The region in the $m_A$--$\tan\beta$ plane where the decay
$H\to hh$ is kinematically {\em forbidden} is indicated by a solid 
line contour.
Also given are contours at which the branching ratio equals 0.1 (dotted),
0.5 (dashed) and 0.9 (dash-dotted, far left).
\end{capt}
\end{center}
\end{figure}

There is a sizeable region in the $m_A$--$\tan\beta$ plane
where the decay $H\to hh$ is kinematically forbidden,
which is shown in Fig.~\ref{Fig:hole-1000}
as an egg-shaped region at the upper left of the plot.
The boundary of the region depends 
crucially on the precise Higgs mass values.
This is illustrated by comparing two cases of mixing parameters
$A$ and $\mu$.
We also display the regions where the $H\to hh$
branching ratio is in the range 0.1--0.9.
Obviously, in the forbidden region, the $\lambda_{Hhh}$ cannot be
determined from resonant production.

\subsection{Double Higgs-strahlung}
As discussed above,
for small and moderate values of $\tan\beta$, a study of decays
of the heavy $CP$-even Higgs boson $H$ provides a means of determining
the triple-Higgs coupling $\lambda_{Hhh}$.
For the purpose of extracting the coupling $\lambda_{hhh}$, 
non-resonant processes
involving two-Higgs ($h$) final states must be considered.
The $Zhh$ final states produced in the non-resonant double 
Higgs-strahlung $e^+e^- \rightarrow Zhh$, and whose cross section
involves the coupling $\lambda_{hhh}$, 
could provide one possible opportunity.
\begin{figure}[htb]
\refstepcounter{figure}
\label{Fig:sig-Zll-2}
\addtocounter{figure}{-1}
\begin{center}
\setlength{\unitlength}{1cm}
\begin{picture}(12,7.8)
\put(0.5,1.5)
{\mbox{\epsfysize=7.0cm\epsffile{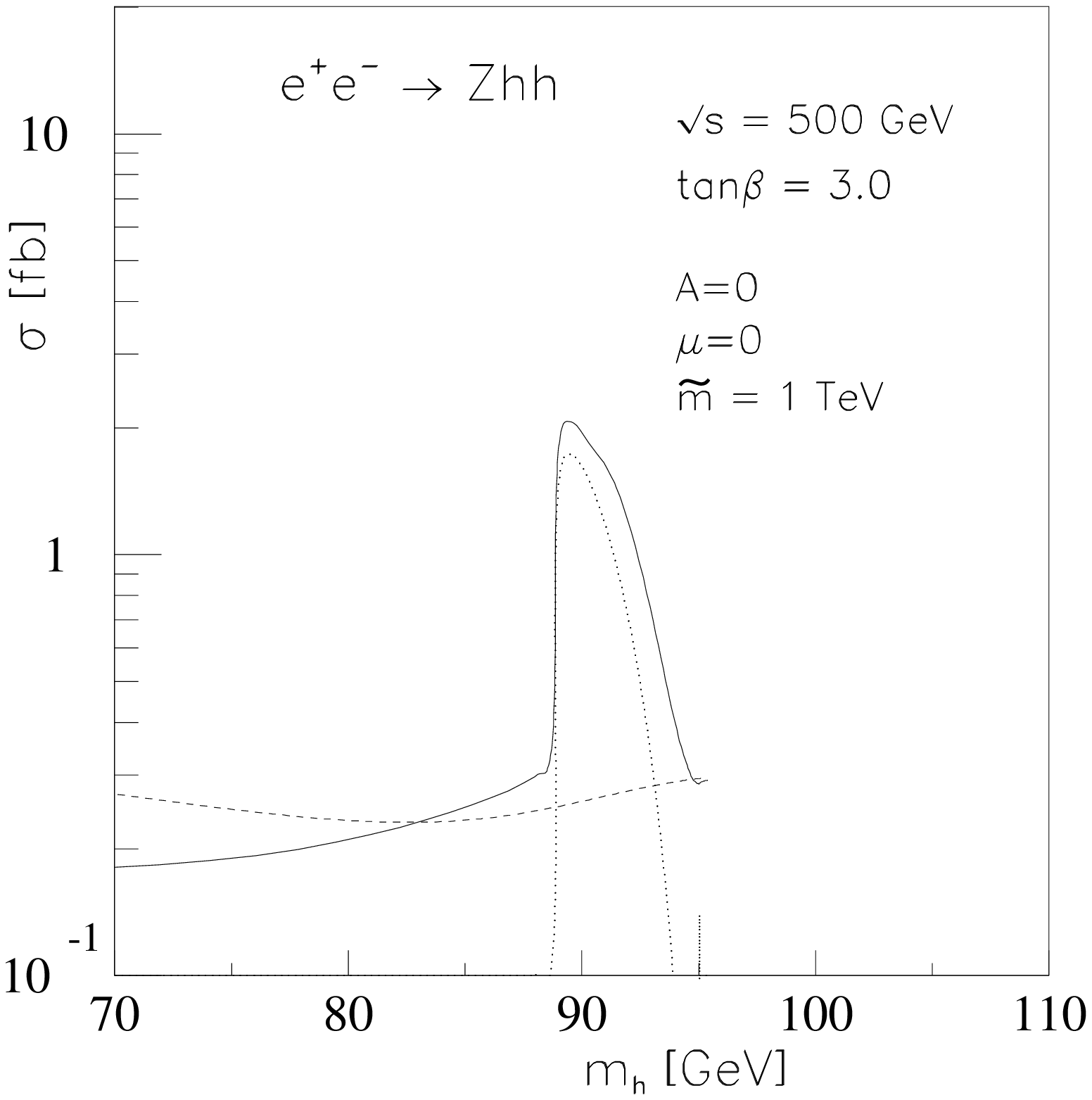}}
 \mbox{\epsfysize=7.0cm\epsffile{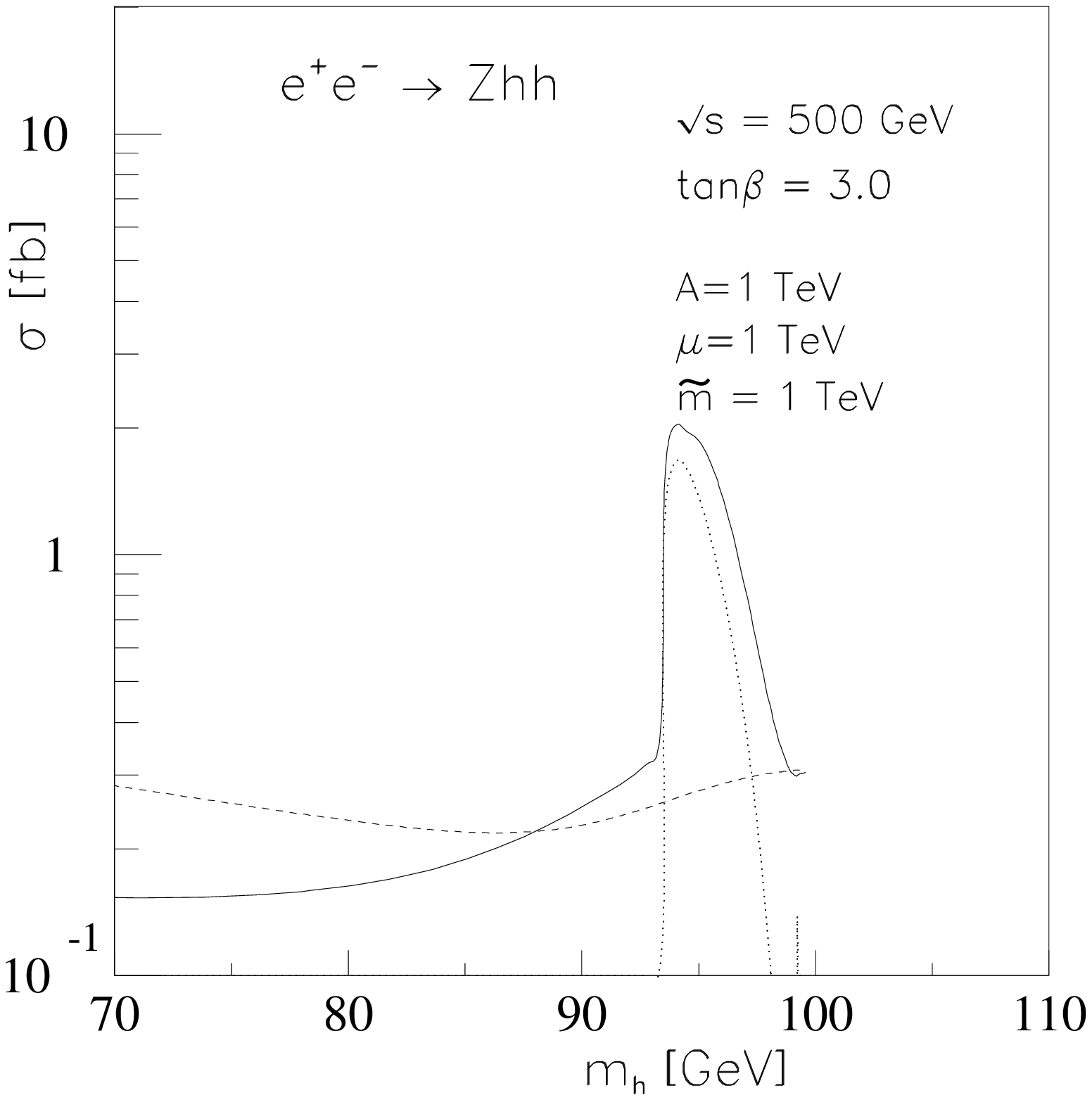}}}
\end{picture}
\vspace*{-22mm}
\begin{capt}
Cross section $\sigma(e^+e^-\to Zhh)$ as a function of $m_h$.
The dotted curve is the resonant production,
the dashed curve gives the decoupling limit \cite{HABER1}.
\end{capt}
\end{center}
\end{figure}

However, the non-resonant contribution to the $Zhh$ cross section
is rather small, as is shown 
in Fig.~\ref{Fig:sig-Zll-2} for $\sqrt s=500~\GeV$, $\tan\beta=3.0$,
and $\tilde m = 1~\TeV$.
At low values of $m_h$, the decay $H\to hh$ is kinematically forbidden.
This is followed by an increase of the trilinear couplings.

Since the non-resonant part of the cross section, which depends on 
$\lambda_{hhh}$, is rather small, this channel is not suitable for
a determination of $\lambda_{hhh}$  \cite{DHZ}.

\subsection{Fusion mechanism for multiple-$h$ production}
A two-Higgs ($hh$) final state
can also result from the $WW$ fusion mechanism in $e^+ e^-$ collisions.
There is a resonant contribution (through $H$) and a non-resonant one. 

The resonant $WW$ fusion cross section for 
$e^+e^- \rightarrow H\bar\nu_e\nu_e$ \cite{DHKMZ}
is plotted in Fig.~\ref{Fig:sigma-500-1500} 
for the centre-of-mass energy $\sqrt s = 500$ GeV, 
and for $\tan\beta = 3.0$, as a function of $m_H$.  

Besides the resonant $WW$ fusion mechanism for the multiple
production of $h$ bosons, there is also a non-resonant $WW$ 
fusion mechanism:
\begin{equation}
\label{Eq:WW-nonres}
e^+e^-\to\nu_e\bar\nu_e hh,
\end{equation}
through which the same final state of two $h$ bosons can be produced.
The cross section for this process (see Fig.~\ref{Fig:Feynman-nonres-WW}),
can be written in the effective $WW$ approximation as
a $WW$ cross section, at invariant energy squared $\hat s=xs$, 
folded with the $WW$ ``luminosity'' \cite{CDCG}. Thus,
\begin{equation}
\label{Eq:sigWW-nonres}
\sigma(e^+e^-\to\nu_e\bar\nu_e hh)
=\int_\tau^1\dd x\, \frac{\dd L}{\dd x}\, \hat\sigma\sup_{WW}(x),
\end{equation}
where $\tau=4m_h^2/s$, and
\begin{equation}
\frac{\dd L(x)}{\dd x}
=\frac{G_{\rm F}^2m_W^4}{2}\,\left(\frac{1}{2\pi^2}\right)^2
\frac{1}{x}\biggl\{(1+x)\log\frac{1}{x}-2(1-x)\biggr\}.
\end{equation}

The $WW$ cross section receives contributions from several amplitudes,
according to the diagrams (a)--(d) 
in Fig.~\ref{Fig:Feynman-nonres-WW}, only one of which
is proportional to $\lambda_{hhh}$.
We have evaluated these contributions \cite{OP98}, 
following the approach of Ref.~\cite{AMP}, ignoring transverse
momenta everywhere except in the $W$ propagators.
Our approach also differs from that of \cite{DHZ} in that 
we do not project out 
the longitudinal degrees of freedom of the intermediate $W$ bosons.

\begin{figure}[htb]
\refstepcounter{figure}
\label{Fig:sig-WW-2}
\addtocounter{figure}{-1}
\begin{center}
\setlength{\unitlength}{1cm}
\begin{picture}(12,7.8)
\put(0.5,1.5)
{\mbox{\epsfysize=7.0cm\epsffile{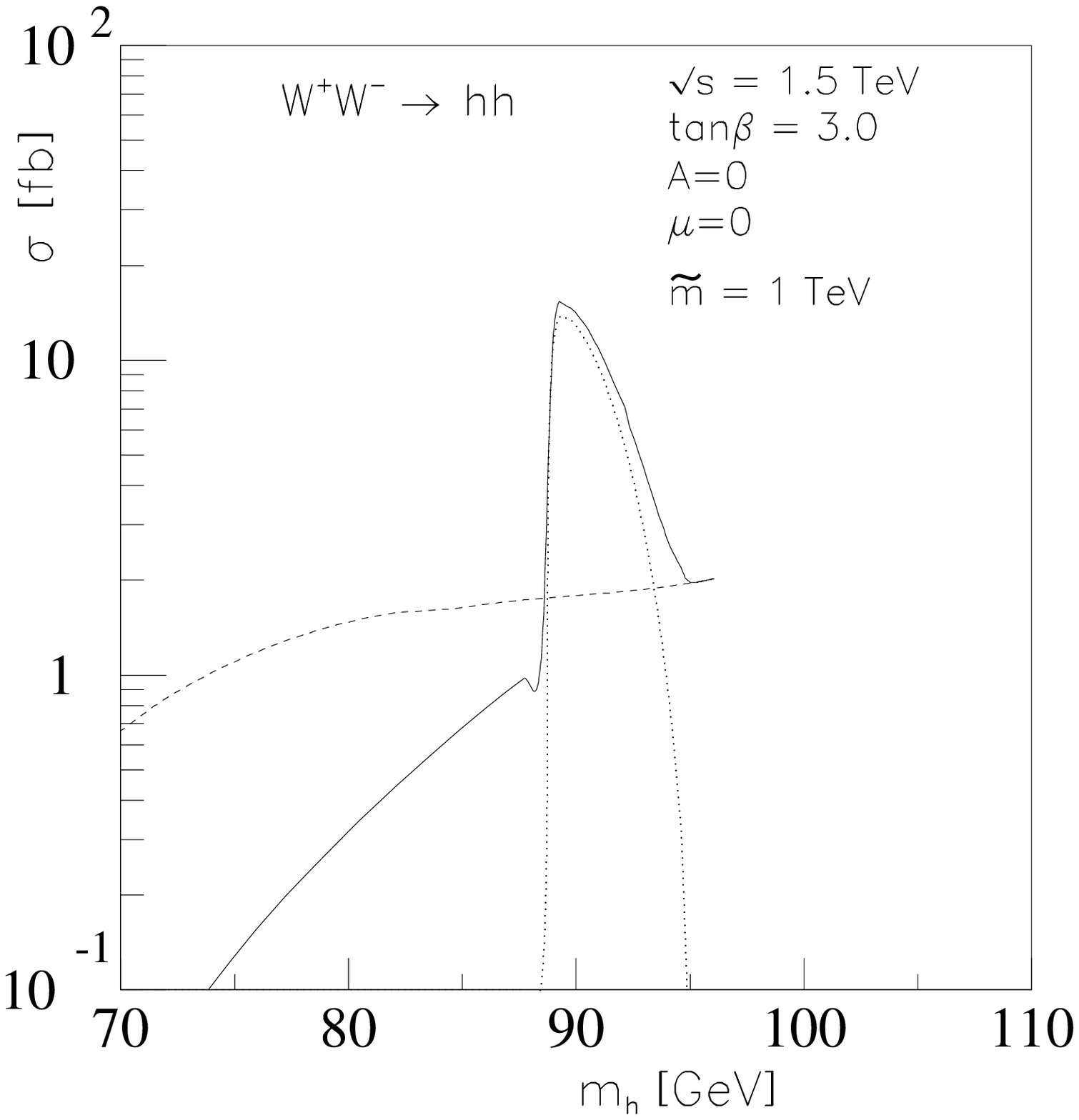}}
 \mbox{\epsfysize=7.0cm\epsffile{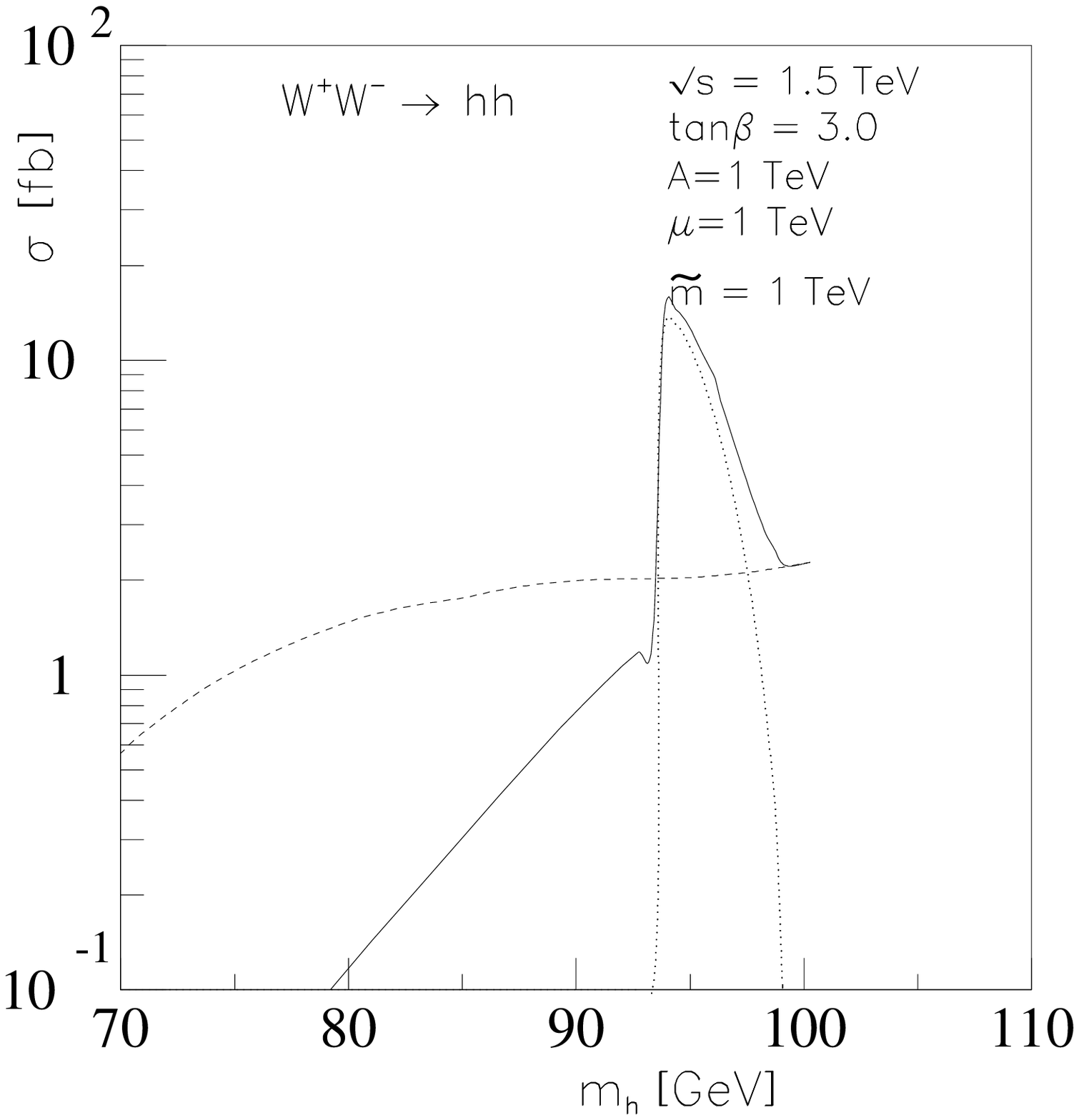}}}
\end{picture}
\vspace*{-22mm}
\begin{capt}
Cross section $\sigma(e^+e^-\to \nu_e\bar\nu_e hh)$ 
(via $WW$ fusion) as a function of $m_h$.
The dotted curve is the resonant production,
the dashed curve gives the decoupling limit.
\end{capt}
\end{center}
\end{figure}

We show in Fig.~\ref{Fig:sig-WW-2} the resulting $WW$ fusion cross section, 
at $\sqrt{s}=1.5~\TeV$, and for $\tilde m = 1~\TeV$.
The structure is reminiscent of Fig.~\ref{Fig:sig-Zll-2},
and the reasons for this are the same. Notice, however, that 
the scale is different. Since this is a fusion cross section,
it grows logarithmically with energy.

For high values of $m_h$ we see that there is a moderate contribution
to the cross section from the non-resonant part.
For a lower squark mass scale $\tilde m$, the situation is 
rather similar, except that the cross section peak (from resonant
production) gets shifted to a higher Higgs mass, $m_h$.
\section{Sensitivity to $\lambda_{Hhh}$ and $\lambda_{hhh}$}
We are now ready to combine the results and discuss in which parts
of the $m_A$--$\tan\beta$ plane one might hope to measure the trilinear
couplings $\lambda_{Hhh}$ and $\lambda_{hhh}$.
In Figs.~\ref{Fig:sensi-500-m1000} and \ref{Fig:sensi-500-m500}
we have identified regions according to the following criteria 
\cite{DHZ,OP98}:
\begin{itemize}
\item[(i)]
Regions where $\lambda_{Hhh}$ might become measurable are identified
as those where 
$\sigma(H)\times\mbox{BR}(H\to hh)> 0.1\mbox{ fb}$ (solid),
while simultaneously $0.1 < \mbox{BR}(H\to hh) < 0.9$
[see Figs.~\ref{Fig:BR-H-A}--\ref{Fig:hole-1000}].
In view of the recent, more optimistic, view on the
luminosity that might become available, 
we also give the corresponding contours for 0.05~fb (dashed) 
and 0.01~fb (dotted). 
\item[(ii)]
Regions where $\lambda_{hhh}$ might become measurable
are those where the {\it continuum} $WW\to hh$
cross section [Eq.~(\ref{Eq:sigWW-nonres})] is larger than 
0.1~fb (solid).
Also shown are contours at 0.05 (dashed) and 0.01~fb (dotted).
\end{itemize}
We have excluded from the plots the region where $m_h<72.2~\GeV$
\cite{ALEPH98}.
This corresponds to low values of $m_A$ and low $\tan\beta$.

\begin{figure}[htb]
\refstepcounter{figure}
\label{Fig:sensi-500-m1000}
\addtocounter{figure}{-1}
\begin{center}
\setlength{\unitlength}{1cm}
\begin{picture}(12,7.8)
\put(0.5,1.5)
{\mbox{\epsfysize=7.0cm\epsffile{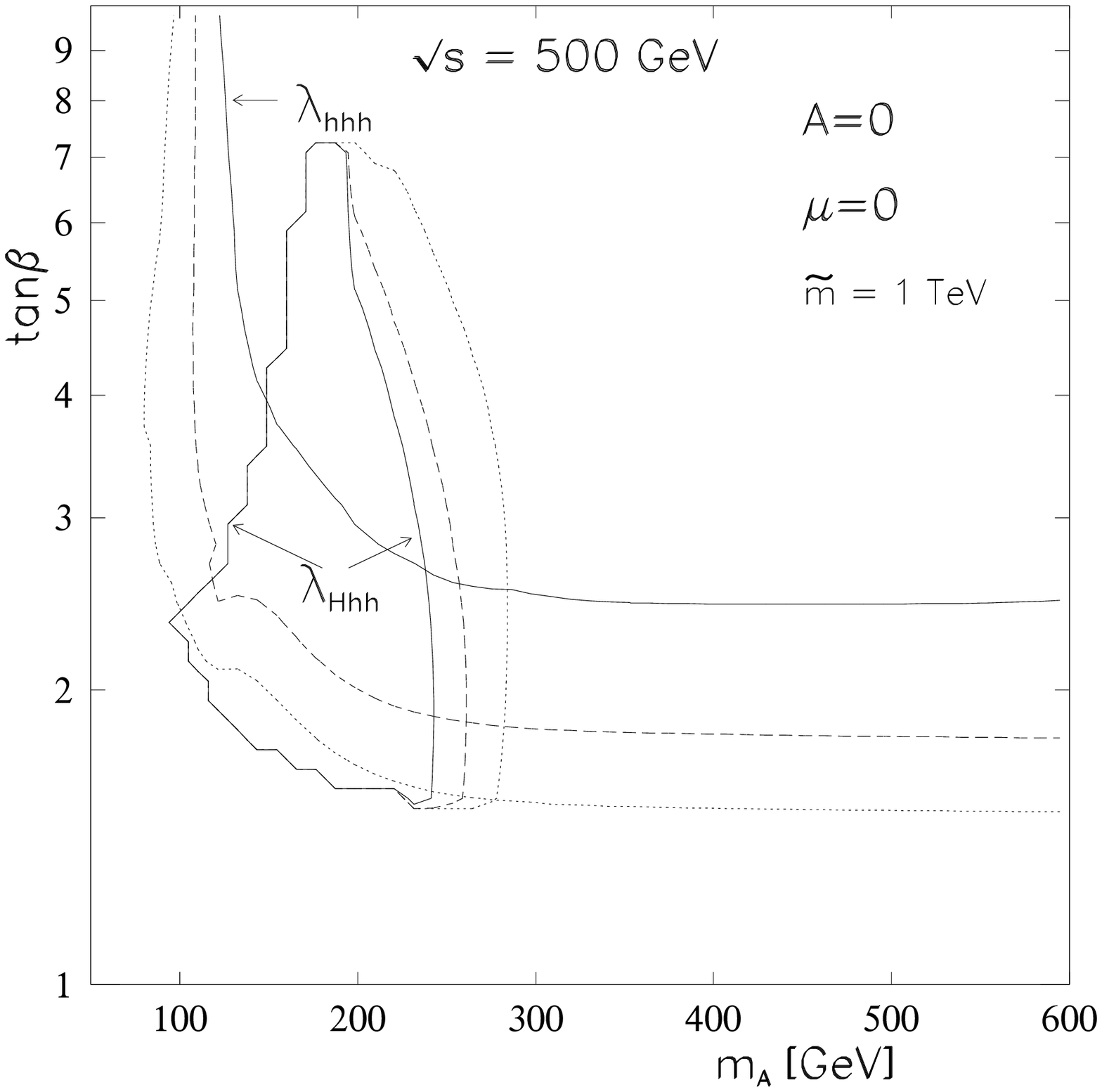}}
 \mbox{\epsfysize=7.0cm\epsffile{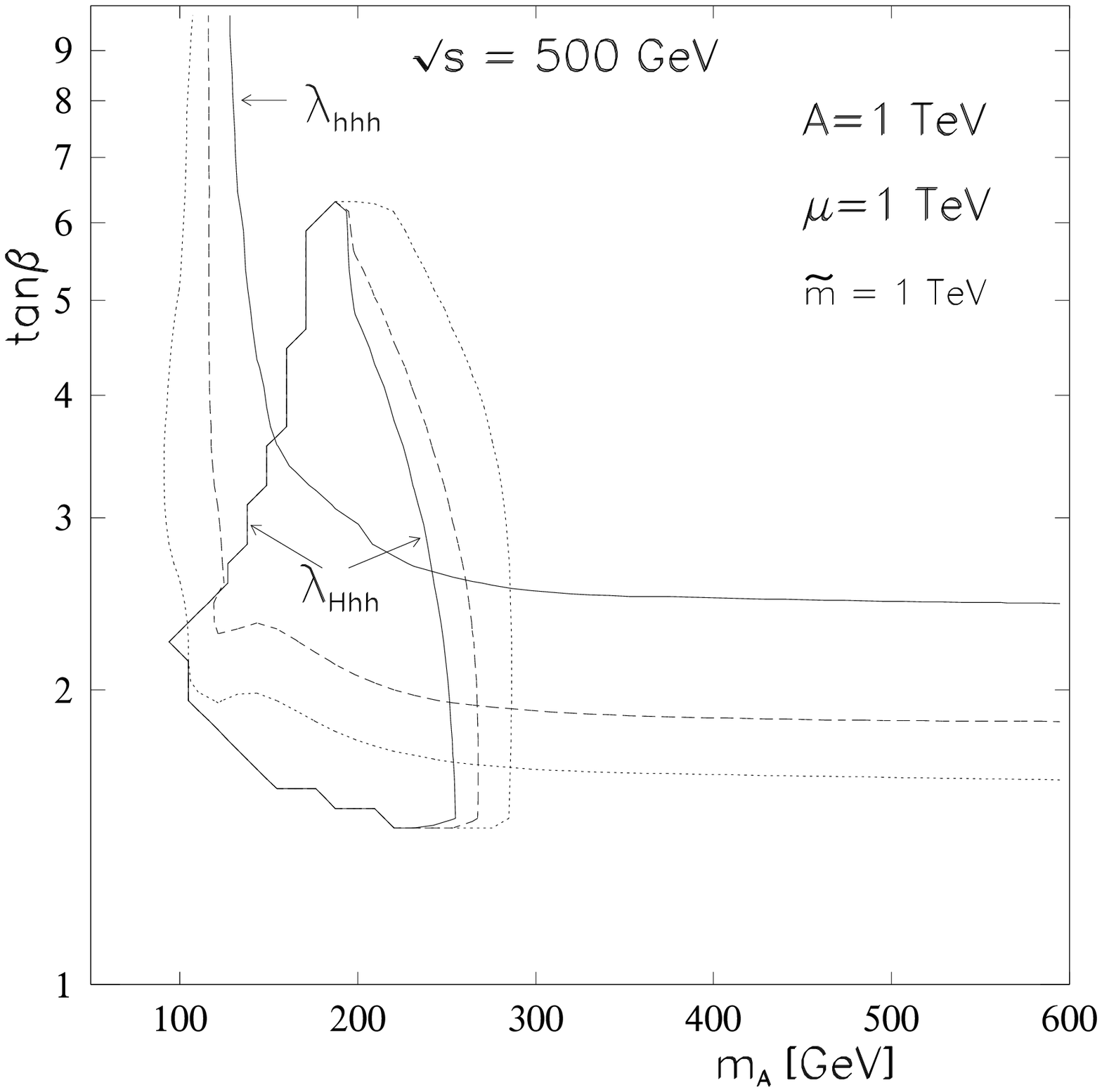}}}
\end{picture}
\vspace*{-22mm}
\begin{capt}
Regions where trilinear couplings $\lambda_{Hhh}$ and 
$\lambda_{hhh}$ might be measurable at $\sqrt{s}=500$~GeV.
Inside contours labelled $\lambda_{Hhh}$, 
$\sigma(H)\times\mbox{ BR}(H\to hh) > 0.1~\mbox{fb}$ (solid),
while $0.1<\mbox{BR}(H\to hh)<0.9$.
Inside (to the right or below) contour labelled $\lambda_{hhh}$,
the {\it continuum} $WW\to hh$ cross section exceeds 0.1~fb (solid).
Analogous contours are given for 0.05 (dashed) and 0.01~fb (dotted).
Two cases of squark mixing are considered, as indicated.
\end{capt}
\end{center}
\end{figure}
\begin{figure}[htb]
\refstepcounter{figure}
\label{Fig:sensi-500-m500}
\addtocounter{figure}{-1}
\begin{center}
\setlength{\unitlength}{1cm}
\begin{picture}(12,7.8)
\put(-1.0,1.5)
{\mbox{\epsfysize=7.0cm\epsffile{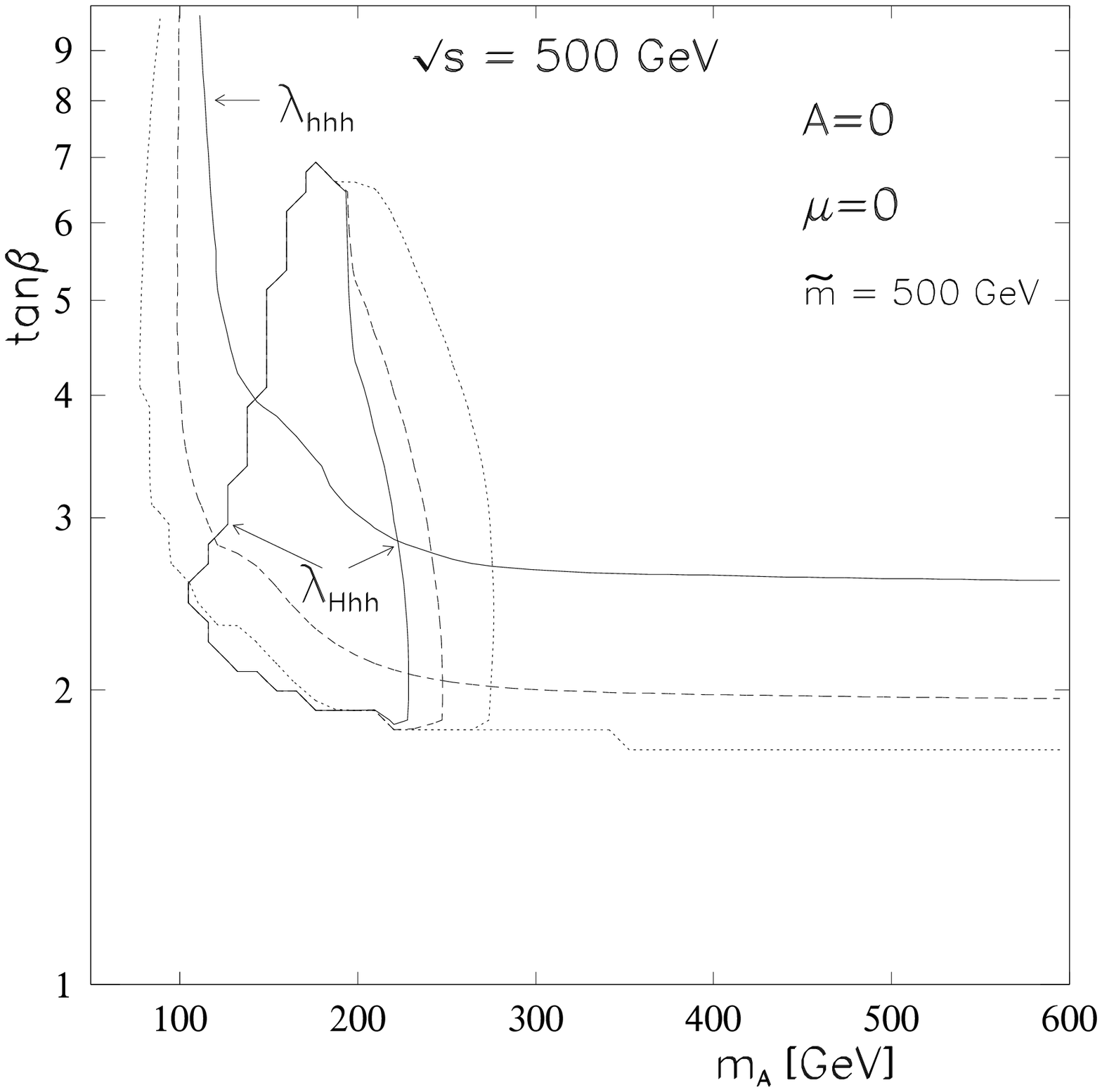}}
 \mbox{\epsfysize=7.0cm\epsffile{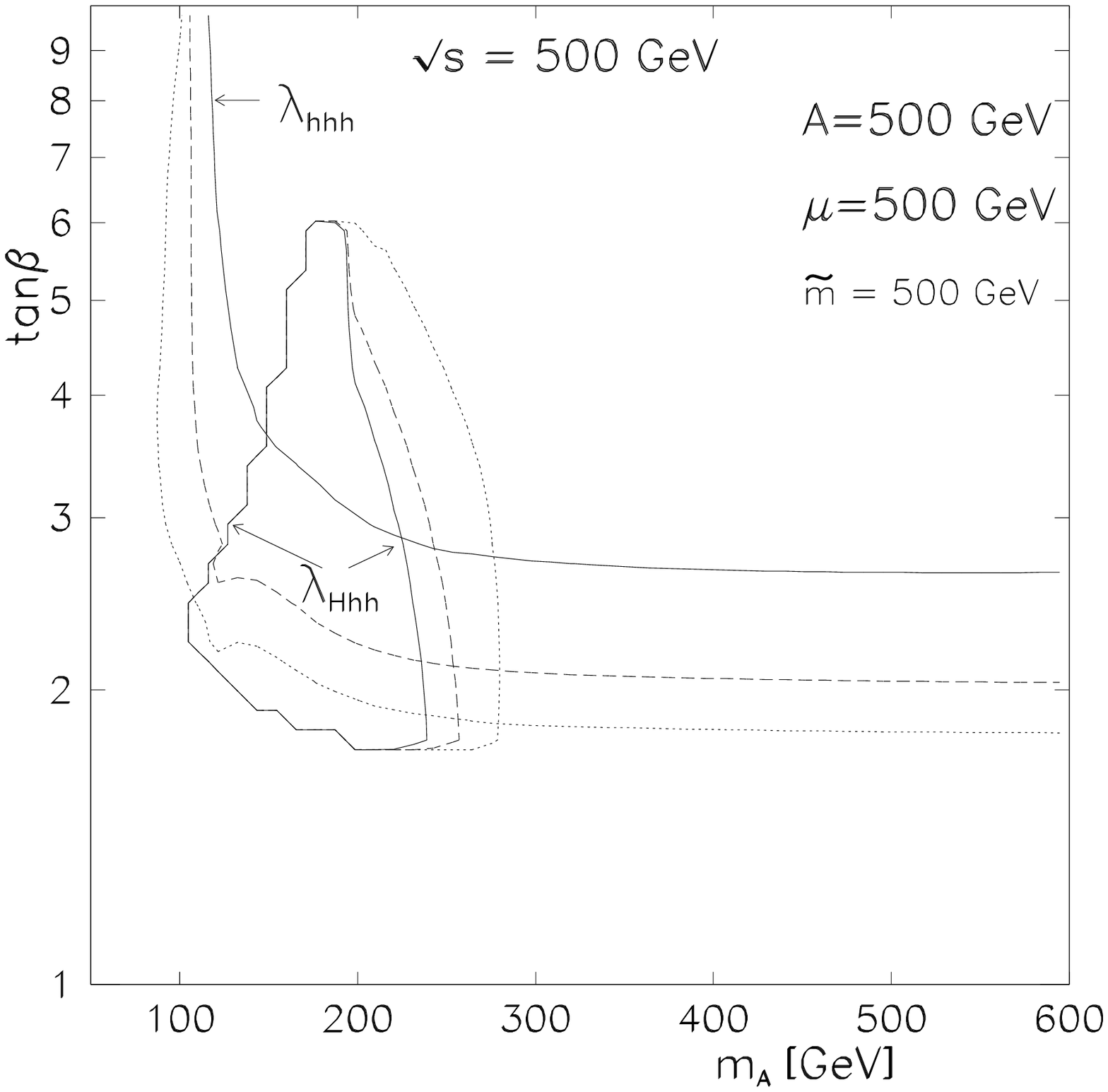}}}
\end{picture}
\vspace*{-22mm}
\begin{capt}
Similar to Fig.~\ref{Fig:sensi-500-m1000} for
$\tilde m=500~\GeV$.
\end{capt}
\end{center}
\end{figure}

These cross sections are small, the measurements are not
going to be easy.
With an integrated luminosity of 500~fb$^{-1}$,
the contours at 0.1~fb correspond to 50 events per year.
This will be reduced by efficiencies, but should indicate
the order of magnitude that can be reached.
For the case of the SM Higgs, the backgrounds have been studied
\cite{background} and the measurements appear feasible.

With increasing luminosity, the region where $\lambda_{Hhh}$ might
be accessible, extends somewhat to higher values of $m_A$.
Note the steep edge around $m_A\simeq200~\GeV$, where increased 
luminosity does not help.
This is determined by the vanishing of $\mbox{BR}(H\to hh)$,
as seen in  Fig.~\ref{Fig:hole-1000}.
The coupling $\lambda_{hhh}$ is accessible in a much larger part
of this parameter space.
 
The precise region in the $\tan\beta$--$m_A$ plane, in which these
couplings might be accessible, depends on details of the model.
As a further illustration of this point, 
we show in Fig.~\ref{Fig:sensi-500-m500}
the corresponding plots for a squark mass parameter $\tilde m=500~\GeV$.

\section{Conclusions}
We have reviewed the updated results of a detailed investigation 
\cite{OP98} of the possibility of measuring
the MSSM trilinear couplings $\lambda_{Hhh}$ and $\lambda_{hhh}$
at an $e^+ e^-$ collider, focussing in detail 
on the importance of mixing in the squark sector,
as induced by the trilinear coupling $A$ and the bilinear coupling $\mu$.
As compared with our earlier work, we here use two-loop results for
the Higgs masses \cite{twoloop}. 
The two-loop results for the Higgs masses have considerably less 
dependence on squark mixing.

As a result of the two-loop Higgs masses being less sensitive
to squark mixing, the regions in the $m_A$--$\tan\beta$ plane 
that are accessible for studying $\lambda_{Hhh}$ and $\lambda_{hhh}$
are more stable than was the case for the one-loop results.

\medskip

This research was supported by the Research Council of Norway,
and (PNP) by the University Grants Commission, India
under project number 10-26/98(SR-I).



\begin{thebibliography}{abc}

\bibitem{Zerwas} P.A. McNamara, Proceedings of the 29th
International Conference on High Energy Physics (ICHEP '98), 
Vancouver 1998,
eds.\ A. Ashtbury, D. Axen and J. Robinson, World Scientific (1999),
p. 1303.
See also P.~M.\ Zerwas, Acta Phys.\ Polon.\ {\bf B30} (1999) 1871.

\bibitem{LEP}
G. Altarelli, R. Barbieri, F. Caravaglios,
Int.\ J. Mod.\ Phys.\ {\bf A13} (1998) 1031.

\bibitem{HPN}  For reviews, see 
H.-P.\ Nilles, Phys.\ Rep.\ {\bf 110}, 1 (1984); 
H.~E.\ Haber and
G.~L.\ Kane, Phys.\ Rep.\ {\bf C117}, 75 (1985);
R. Barbieri, Riv.\ Nuovo Cimento {\bf 11} No.\ 4, p.~1 (1988).

\bibitem{GHKD} J.~F. Gunion, H.~E. Haber, G. Kane and S. Dawson,
{\em The Higgs Hunter's Guide}, Addison-Wesley, New York, 1990.

\bibitem{DHZ} A. Djouadi, H.~E.\ Haber and P.~M.\ Zerwas, 
Phys.\ Lett.\ {\bf B375}, 203 (1996) and Erratum, to be published.

\bibitem{OP98} P. Osland and P. N. Pandita, 
Phys.\ Rev.\ {\bf D~59} (1999) 055013;
Invited paper, 
{\it VIIIth UNESCO St.~Petersburg International School of Physics},
May 25 -- June 4, 1998 (to be published in the Proceedings)
(Archive: hep-ph/9902270);
P. Osland, Acta Phys.\ Polon.\ {\bf B30} (1999) 1967.

\bibitem{Muhlleitner}
A. Djouadi, W. Kilian, M. Muhlleitner and P. M. Zerwas, 
Eur.\ Phys.\ J. {\bf C10} (1999) 45.

\bibitem{NLC}
M. Tigner, B. Wiik and F. Willeke, 
Particle Accel.\ Conf.\ IEEE (1991) 2910; 
S. Kuhlman {\it et al.}
(The NLC ZDR Design Group and the NLC Physics Working Groups),
{\it Physics and Technology of the Next Linear Collider}: 
A Report Submitted to Snowmass '96,
BNL 52-502, FERMILAB-PUB-96/112, LBNL-PUB-5425, SLAC-Report-485,
UCRL-ID-124160, UC-414 (June 1996); 
H. Murayama and M. E. Peskin, Ann.\ Rev.\ Nucl.\ Part.\ Sci.\
{\bf 46} (1996) 533; 
e+e$-$ Linear Colliders: Physics and Detector Studies,
R. Settles, editor, 
Proceedings, workshops, ECFA/DESY, Frascati, London, Munich, Hamburg,
DESY-97-123E, 1997.

\bibitem{ERZ1} J. Ellis, G. Ridolfi and F. Zwirner, 
Phys.\ Lett.\ {\bf B257}, 83 (1991);
Y. Okada, 
M.~Yamaguchi and T. Yanagida, 
Prog.\ Theor.\ Phys.\ {\bf 85}, 1 (1991);
H.~E. Haber and R. Hempfling, Phys.\ Rev.\ Lett.\ 
{\bf 66}, 1815 (1991);
J. Ellis, G. Ridolfi and F. Zwirner, 
Phys.\ Lett.\ {\bf B262}, 477 (1991); 
R. Hempfling and A. H. Hoang, Phys.\ Lett.\ {\bf B331}, 99 (1994); 
M. Carena, J.R. Espinosa, M. Quir\'os and C.E.M. Wagner,
Phys.\ Lett.\ {\bf B355}, 209 (1995); 
M. Carena, M. Quir\'os and C.E.M. Wagner,
Nucl.\ Phys.\ {\bf B461}, 407 (1996); 
S. Heinemeyer, W. Hollik and G. Weiglein, Phys.\ Rev.\ {\bf D58},
091701 (1998).

\bibitem{BBSP} V. Barger, M.~S. Berger, A.~L. Stange 
and R.~J.~N. Phillips, Phys.\ Rev.\ {\bf D45}, 4128 (1992).

\bibitem{twoloop}
S. Heinemeyer, W. Hollik and G. Weiglein, 
Eur.\ Phys.\ J. {\bf C9} (1999) 343; hep-ph/9812320;
Phys.\ Lett.\ {\bf B455} (1999) 179.

\bibitem{ALEPH98}
R. Barate {\it et al.} (ALEPH Collaboration), 
Phys.\ Lett.\ {\bf B} 440 (1998) 419.

\bibitem{PocZsi}
G. P\'ocsik and G. Zsigmond, Z. Phys.\ {\bf C10}, 367 (1981).

\bibitem{GETAL} J. F. Gunion, L. Roszkowski, A. Turski, H. E. Haber, 
G. Gamberini, B. Kayser, S. F. Novaes, F. Olness and J. Wudka, Phys. Rev.
{\bf D38}, 3444 (1988).

\bibitem{DKZ1} For a review, see
A. Djouadi, J. Kalinowski and P. M. Zerwas, 
Z. Phys.\ {\bf C70}, 435 (1996).

\bibitem{HABER1} H. E. Haber, in
Proceedings of the Conference on
{\it Perspectives for Electroweak Interactions in e$^+$ e$^-$ Collisions},
Ringberg (Tegernsee), Germany,
1995; ed. B. A. Kniehl (World Scientific, Singapore, 1995) p.~219.

\bibitem{DHKMZ} A. Djouadi, D. Haidt, B. A. Kniehl, B. Mele and 
P. M. Zerwas, 
in Proceedings, Workshop on
{\it $e^+e^-$ Collisions at 500 GeV: The
Physics Potential}, Munich-Annecy-Hamburg (DESY 92-123~A,
Hamburg, 1992).

\bibitem{CDCG} R. N. Cahn and S. Dawson, Phys. Lett. {\bf B136}, 196 (1984); 
S. Dawson, Nucl.\ Phys. {\bf B249}, 42 (1984);
M. Chanowitz and M. K. Gaillard, Phys.\ Lett.\ {\bf B142}, 85 (1984);
I. Kuss and H. Spiesberger, Phys.\ Rev.\ {\bf D53}, 6078 (1996).

\bibitem{AMP} G. Altarelli, B. Mele and F. Pitolli,
Nucl.\ Phys.\ {\bf B287}, 205 (1987).

\bibitem{background}
D.J. Miller and S. Moretti, RAL-TR-1999-032 (May 1999),
hep-ph/9906395;
P. Lutz, talk at ECFA/DESY Linear Collider Workshop, Oxford, March 1999.

\end{thebibliography}
\end{document}